\documentclass[prb,twocolumn,showpacs,floatfix,amsfonts]{revtex4}
\usepackage{graphicx,graphics,color,epsfig}
\usepackage{rotating}
\usepackage{multirow}
\usepackage{longtable}
\usepackage{bm}
\usepackage{amsmath}
\usepackage{amssymb}
\usepackage{epstopdf}
\begin{document}
\title{Renormalization Group for Mixed Fermion-Boson Systems}
\author{Seiji J. Yamamoto and Qimiao Si}
\affiliation{Department of Physics \& Astronomy, Rice University,
Houston, TX 77005, USA }
\begin{abstract}
We formulate a momentum-shell renormalization group (RG) procedure 
that can be used in theories containing both bosons and fermions
with a Fermi surface. We focus on boson-fermion couplings that 
are nearly forward-scattering, {\it i.e.} involving
small momentum transfer ($\vec{q} \approx 0$) for the fermions. 
Special consideration is given to phase space constraints 
that result from the conservation of momentum and the imposition
of ultraviolet cutoffs.
For problems where the energy and momentum scale similarly
(dynamic exponent $z = 1$), we show that more than one 
formalism can be used and they give equivalent results.
When the energy and momentum must scale differently ($z \neq 1$),
the procedures available are more limited but a consistent RG scheme
can still be formulated.
The approach 
is applicable to a variety of problems,
such as itinerant-electron magnets and gauge fields
interacting with fermions.
\end{abstract}
\pacs{05.10.Cc, 64.60.ae}
\maketitle

\section{Introduction}
\label{sec:intro}
Although the theory of scaling and renormalization has profoundly
affected our conceptual understanding of many-body systems, its calculational
framework is imperfect and continually evolving. In the end, we are interested
in how couplings flow under changes of scale, but a variety of distinct
procedures exist, each with its own advantages and drawbacks.   An incomplete
list of the assortment of programs includes the multiplicative RG, real space
decimation, functional RG, exact RG, flow equations, and various flavors of 
$\epsilon$-expansion, such as the classic minimal subtraction
which expands around $d=4$, or expansions around some other parameter,
such as the deviation of the range of the interaction from a suitable 
reference value. Each method has its own limits of practicality, ease of use,
and range of problems to which it may be usefully employed.  
Wilson's momentum-shell approach~\cite{Wilson1974, Hertz1976} 
is an especially popular method in the context of condensed matter problems.  
However, in the early 1990's a few people
recognized~\cite{Feldman1990, Benfatto1990, Shankar1991} that the standard 
momentum-shell procedure must be modified for problems involving a Fermi surface.
A campaign soon followed attempting to understand Fermi liquid theory from 
an RG perspective.  An excellent and influential summary of the pure fermion
RG can be found in~\cite{Shankar1994}.

Another indication that the RG for fermions required more scrutiny came from
the study of  quantum criticality in itinerant electron magnets.  The usual
Hertzian approach~\cite{Hertz1976} uses an auxiliary (Hubbard-Stratonovich) 
field to decouple the fermion-fermion interaction,
thus allowing
fermions to be completely integrated out.  
The resulting effective theory is then expressed in terms of 
the remaining bosonic auxiliary field, to which standard bosonic RG techniques
can be employed.  
However, because the fermions are gapless,
the process of integrating them out may 
introduce non-analyticities in the couplings
among the remaining bosonic 
modes~\cite{Vojta1997, Abanov2004}.
It would therefore
be important 
to devise an RG scheme capable
of simultaneously handling both bosons and fermions with a Fermi surface.

Besides the critical itinerant magnets, a mixed fermionic-bosonic
RG formalism would be quite useful for an assortment of problems. 
For example, in the context of a gauge field coupled to fermions, several
authors~\cite{Polchinski1994, Nayak1994, Altshuler1994, Onoda1995} have developed
their own RG schemes for counting dimensions in these mixed theories.  All have in
common the subdivision of the Fermi surface into a large number of patches, 
but results vary and despite the intervening 15 years since this pioneering work,
little progress has been made.  
The importance of the gauge-fermion problem is historically linked to an 
interesting path to non-Fermi liquid behavior \cite{Holstein1973, Varma2002}.
More recently, effective gauge theories have appeared in a number of 
additional contexts in 
condensed matter physics \cite{Lee2006}.

We should mention in passing a growing body of work on the functional RG which
may be adapted for mixed boson-fermion
theories~\cite{Furukawa1998, Zanchi2000a, Zanchi2000b, Honerkamp2001, Schutz2005, Tsai2005}.  
This typically requires blending with computational methods and may prove to be a useful
framework for understanding realistic material bandstructures.
Our aim here is rather more modest, which is to develop an RG scheme for mixed theories
with a high score in the ``ease of use'' category.  This was the chief virtue of the 
original Wilsonian RG which could quickly identify the relevant and irrelevant 
operators with minimal effort.
One emphasis of this paper will be to carefully
consider how to extend Shankar's scheme \cite{Shankar1994} to include bosons while
maintaining the easy-to-use spirit of the Wilsonian approach.

Our primary motivation to consider these issues came from the context of magnetically
ordered phases of some itinerant systems. In the case of an antiferromagnetic
state of a Kondo lattice, the bosonic magnons,
described by a quantum non-linear sigma model, are coupled to the fermionic
quasiparticles near a Fermi surface. In this problem, energy and momentum 
scale the same way; the dynamic exponent $z=1$. The ferromagnetic counterpart
features $z=3$. The RG analysis plays an essential role in understanding the
Fermi surfaces in these systems and has been briefly described in our
earlier works~\cite{Yamamoto2007,Yamamoto2008,Yamamoto2008b}. The purpose
of the present work is to explain the details of the method in
considerable detail with the hope that the method will be adapted to problems 
in new physical contexts.

The remainder of the paper is organized as follows. In section~\ref{sec:bosons} we remind the reader 
of the essential points of the bosonic Wilson-Hertz scaling.  Section~\ref{sec:fermions} quickly moves 
on to discuss scaling in fermionic systems, largely paraphrasing what has already been done but 
emphasizing a slightly different perspective.  The next section, \ref{sec:bosons+fermions},
describes a method to properly scale in mixed theories when energy and momentum can be given
the same scaling dimension, \textit{i.e.} when $z=1$.  This is closest in spirit to the Shankar
approach, but cannot be generalized to $z \neq 1$.  Some of these problems are discussed in 
sections \ref{sec:choiceOfScheme} and \ref{sec:patching}.  In section \ref{sec:patching} 
we present an alternative method for arbitrary $z$ that is perhaps less intuitive than 
that of section \ref{sec:bosons+fermions}, but has the advantage of being generalizable
to $z \neq 1$ while at the same time yielding identical results when $z=1$.

\section{Boson Scaling}
\label{sec:bosons}

The problem we are concerned with can be decomposed into bosonic, fermionic, and interaction terms:
\begin{eqnarray}
	\mathcal{S} &=& \mathcal{S}^f + \mathcal{S}^b + \mathcal{S}_3^{bf}
\end{eqnarray}
The bosonic and fermionic pieces can be further divided into quadratic and quartic pieces.
\begin{eqnarray}
	\mathcal{S}^b &=& \mathcal{S}^b_2 + \mathcal{S}^b_4 \\
	\mathcal{S}^f &=& \mathcal{S}^f_2 + \mathcal{S}^f_4
\end{eqnarray}
Theories based upon $\mathcal{S}^b$ or $\mathcal{S}^f$ alone have already been 
subjected to momentum-shell RG analyses; see, for example, 
\cite{Wilson1974, Hertz1976} and \cite{Shankar1994}.
In this section, we review the Wilson-Hertz scaling procedure for bosons, so we are
only concerned with $\mathcal{S}^b$.

In the most general case, the quadratic part of the action can take 
several different forms depending on the value of $z$.  For example,
\begin{eqnarray}
	\mathcal{S}_2^b(z=1) &=& \int d^dq d\omega\; \phi^*(q^2 + \omega^2)\phi 
\nonumber 
\\
	\mathcal{S}_2^b(z=2) &=& \int d^dq d\omega\; \phi^*(q^2 + \omega)\phi \nonumber \\
	\mathcal{S}_2^b(z=3) &=& \int d^dq d\omega\; \phi^* \left(q^2 
+ \frac{\omega}{q} \right)\phi \nonumber \\
	\mathcal{S}_2^b(z=4) &=& \int d^dq d\omega\; \phi^* \left(q^2 +  \frac{\omega}{q^2} \right)\phi 
\end{eqnarray}
The bosons might represent acoustic phonons, magnons, photons, or some 
collective mode of an underlying fermionic theory that results after ``integrating out''
the fermions with an auxiliary field.  
At this point, we need not be specific. 
All that matters is that we must design the RG scheme in such a way that
$\mathcal{S}_2^b$ remains invariant.
The Wilsonian RG for bosons is well-known~\cite{Wilson1974} so we only review 
those elements crucial to the comparisons we wish to make later with the fermionic RG.  

Consider a $d$-dimensional integral in momentum space with a cutoff to high-energy 
and therefore large-$q$ modes; this is denoted by $\Lambda$.  
Let us separate out a thin shell of high energy
modes in the range $\Lambda/s < q < \Lambda$, where $s \gtrapprox 1$.
\begin{eqnarray}
	\int^{\Lambda} d^dq 
	&\equiv&
	\int d^{d-1}\Omega_{\vec{q}} \int_{0}^{\Lambda}q^{d-1}dq  \nonumber \\
	&=&
	\int d^{d-1}\Omega_{\vec{q}} \left[\int_{0}^{\Lambda/s}q^{d-1}dq  + \int_{\Lambda/s}^{\Lambda} q^{d-1}dq \right] \nonumber
\end{eqnarray}
Here, $q \equiv |\vec{q}|$ is the radial coordinate in (hyper)spherical coordinates, 
and $d^{d-1}\Omega_{\vec{q}}$ represents the measure for integration over all angular variables in $\vec{q}$-space.  
We have ignored factors of $2\pi$.  
Mode elimination amounts to simply throwing away the shell integral.  
To regain the original form of the action only a trivial rescaling of the radial coordinate is needed: 
\begin{eqnarray}
	q^{\prime} &\equiv& s q
\label{eq:ScalingQ}
\end{eqnarray}
This defines the scaling dimension of momentum.
In the customary notation,
we use square brackets to denote the 
scaling
dimension of any quantity
according to
\begin{eqnarray}
	A^{\prime} &=& s^{ [A] } A
\label{eq:bracketNotation}
\end{eqnarray}
where $A^{\prime}$ is measured in units $s^{ [A] }$
times smaller than the units of $A$.
We call $[A]$ the scaling dimension of $A$.
This notation differs from another frequent convention which may claim,
for instance, $[\text{volume}] = L^3$ where $L$ is some length scale.
We prefer our notation since it means a coupling g
is relevant when $[g]>0$,
irrelevant when $[g]<0$,
and marginal when $[g]=0$.

In this notation, an equivalent statement to equation $(\ref{eq:ScalingQ})$ is simply
\begin{eqnarray}
	\left[ q \right] &=& 1
\label{eq:qdim}
\end{eqnarray}
Using this form of momentum scaling in the integral leads to
\begin{eqnarray}
	\int d^{d-1}\Omega_{\vec{q}} \int_{0}^{\Lambda} s^{-(d-1)}q^{\prime d-1}s^{-1}\,dq^{\prime} 
	&=&
	s^{-d}\int^{\Lambda} (d^dq)^{\prime} \nonumber
\end{eqnarray}
We conclude that the scaling dimension of the measure is given by
\begin{eqnarray}
	[d^dq] &=& d[q] = d
\label{eq:BosonMeasure}
\end{eqnarray}

Note that rescaling the radial variable, $q$, is the same as rescaling all the components 
of $\vec{q}$ since $q = \sqrt{\sum_{\alpha}^d q_{\alpha}^{ 2} }$.  
For this to be consistent with $q^{\prime} = sq$, we must have $q_{\alpha}^{\prime} = s q_{\alpha}$ 
for all components $\alpha \in \{ x,y,z,...\}$.  
This is an important difference from the fermionic case to be discussed later and
results from the simple fact that the coordinate origin here is a single point rather
than an extended surface.

Let us apply this mode elimination and rescaling to the quadratic part of the boson action, 
taking the case $z=3$ as an example.
\begin{eqnarray}
	\mathcal{S}_2^b(z=3) 
	&=& \int^{\Lambda/s} d^dq d\omega |\phi|^2 \left( q^2 + \frac{\omega}{q} \right) + \text{(high energy)} \nonumber \\
	&\approx& s^{-d[q]-[\omega] } \int^{\Lambda} d^dq^{\prime} d\omega^{\prime} 
	|\phi(s^{-[q]}\vec{q}~^{\prime},s^{-[\omega] } i\omega^{\prime} )|^2 \nonumber \\
	&&\times \left(s^{-2[q] } q^{\prime 2} + s^{-[\omega]+[q]} \frac{\omega^{\prime} }{ q^{\prime} } \right)
\end{eqnarray}
where we discarded high energy modes and used equation (\ref{eq:ScalingQ}) 
and $\omega^{\prime} \equiv s^{[\omega]}\omega$.
We wish to scale the terms in the parentheses identically, so we must have $-2[q] = -[\omega]+[q]$ thus
fixing the relationship between the scaling dimensions of energy and momentum:
\begin{eqnarray}
	\left[ q \right] &=& \left[ \omega \right] / z
\label{eq:Bosonwqz}
\end{eqnarray}
where our example considered $z=3$ explicitly.

The final step is wavefunction renormalization
which can be implemented by defining a new field $\phi^{\prime}$
according to:
\begin{eqnarray}
	\phi^{\prime}(q^{\prime},i\omega^{\prime}) \equiv s^{-\{(d+2)[q]+[\omega] \} /2} \phi(s^{-[q]} q^{\prime}, s^{-[\omega]} i \omega^{\prime})
\label{eq:BosonScaling}
\end{eqnarray}
Equivalently, we might say that the boson field has a scaling dimension
given by
\begin{eqnarray}
	\left[\phi\right] = -\frac{[q]}{2}(d+z+2)
\label{eq:phiDim}
\end{eqnarray}
where we used $[\omega] = z[q]$. 
Although customary, equation (\ref{eq:phiDim}) is slightly misleading. 
The replacement of $\phi$ by $s^{-[\phi]}\phi^{\prime}$ in analyzing interaction
terms should only be done when
the arguments of the field transform according to equation (\ref{eq:BosonScaling}).
Equation (\ref{eq:S3K1K2Rep}) provides an example where the arguments
of the field are transformed in a very different fashion.
From equation (\ref{eq:BosonScaling}) we see that the boson field appears to
take the form of a generalized homogeneous function.  We do not
delve into this issue further, but merely note that equation (\ref{eq:BosonScaling})
is a very specific type of substitution that needs to be implemented in this strict form.
Scale invariance of $\mathcal{S}_2^b$ has imposed a transformation property
on the field, specified in $(\ref{eq:BosonScaling})$, 
under the particular coordinate transformation $q^{\prime} = s^{[q]}q$
and $\omega^{\prime} = s^{[\omega]}\omega$ with $[q]=[\omega]/z$.

Now that we know how to scale 
momentum from equation (\ref{eq:qdim}),
energy from equation (\ref{eq:Bosonwqz}),
and the field from equation (\ref{eq:BosonScaling}),
we are ready to analyze the four-boson interaction term:
\begin{eqnarray}
	\mathcal{S}_4^b 
	&=& u_b \int 
		d^dq_4 d\omega_4
		d^dq_3 d\omega_3
		d^dq_2 d\omega_2
		d^dq_1 d\omega_1 \nonumber\\
&&\times	\phi(\vec{q}_4,i\omega_4)
		\phi(\vec{q}_3,i\omega_3)
		\phi(\vec{q}_2,i\omega_2)
		\phi(\vec{q}_1,i\omega_1) \nonumber \\
&&\times	\Theta(\Lambda - |\vec{q}_4|)
		\Theta(\Lambda - |\vec{q}_3|)
		\Theta(\Lambda - |\vec{q}_2|)
		\Theta(\Lambda - |\vec{q}_1|) \nonumber \\
&&\times	\delta^{(d)}(\vec{q}_4+\vec{q}_3 - \vec{q}_2 - \vec{q}_1)
		\delta(\omega_4+\omega_3 - \omega_2 - \omega_1) \nonumber \\
\end{eqnarray}
The $\delta$-functions enforce the conservation of energy and momentum while
the $\Theta$-functions define the cutoffs for the effective field theory 
(in principle, energy cutoffs should also be written, but this is understood).

To determine the scaling dimension of $u_b$ at the tree level, we first
separate the integrations into low and high energy modes 
[\textit{i.e.} 
$\Theta(\Lambda - |\vec{q}_i|)
=
\Theta(\Lambda/s - |\vec{q}_i|) +
\Theta(|\vec{q}_i| - \Lambda/s)\Theta(\Lambda - |\vec{q}_i|)
$], 
then discard the high energy shell.
There is some freedom in choosing the shape of the shell
which can take some curious forms for the purpose of
simplifying calculations.
See the discussion by Hertz~\cite{Hertz1976}.  

After rescaling according
to (\ref{eq:qdim}), (\ref{eq:Bosonwqz}), and (\ref{eq:phiDim}), we find
\begin{eqnarray}
	\mathcal{S}_4^b 
	&=& s^{4-d-z} u_b \int 
		d^dq_3^{\prime} d\omega_3^{\prime}
		d^dq_2^{\prime} d\omega_2^{\prime}
		d^dq_1^{\prime} d\omega_1^{\prime} \nonumber\\
&&\times	\phi(\vec{q}_4^{\prime},i\omega_4^{\prime})
		\phi(\vec{q}_3^{\prime},i\omega_3^{\prime})
		\phi(\vec{q}_2^{\prime},i\omega_2^{\prime})
		\phi(\vec{q}_1^{\prime},i\omega_1^{\prime}) \nonumber \\
&&\times	\Theta(\Lambda - |\vec{q}_4^{\prime}|)
		\Theta(\Lambda - |\vec{q}_3^{\prime}|)
		\Theta(\Lambda - |\vec{q}_2^{\prime}|)
		\Theta(\Lambda - |\vec{q}_1^{\prime}|) \nonumber \\
&&\times	\delta^{(d)} \left( \vec{q}_4^{\prime}+\vec{q}_3^{\prime} - \vec{q}_2^{\prime} - \vec{q}_1^{\prime} \right)
		\delta \left( \omega_4^{\prime}+\omega_3^{\prime} - \omega_2^{\prime} - \omega_1^{\prime} \right) \nonumber \\
\end{eqnarray}
which tells us that $u_b^{\prime} \equiv s^{4-d-z} u_b$,
or equivalently
\begin{eqnarray}
	\left[ u_b \right] &=& 4-(d+z)
\end{eqnarray}

This yields a quick way to determine when the four-boson interaction term $u_b \int \phi^4$
is relevant or irrelevant based on the dimensionality of the problem and the value of $z$.  
Historically, this result provided some early intuition
about quantum phase transitions which can behave like classical phase transitions
but in a different number of effective dimensions: $d_{\text{eff}} = d+z$.
Although the theory was originally devised to address questions about
itinerant quantum critical magnets~\cite{Hertz1976, Moriya1985, Millis1993},
some problems have been encountered with this approach~\cite{Vojta1997, Abanov2004}.
Part of the problem could be that the theory is completely bosonic, despite the underlying
fermionic nature of the system.  It is therefore desirable to develop an RG formalism
that includes fermions with a Fermi surface.

\section{Fermion Scaling: Shankar's RG}
\label{sec:fermions}
For fermions, the quadratic part of the action is given by
\begin{eqnarray}
	\mathcal{S}_2^f &=& \int d^dK d\epsilon \; \bar{\psi}\left( i\epsilon - \xi_{\vec{K} } \right)\psi
\end{eqnarray}
To define a scaling scheme that leaves $\mathcal{S}_2^f$ scale invariant, 
we now review the formulation of the fermionic RG~\cite{Shankar1994}.
We shall use Shankar's notation and label momenta measured with respect to the 
Brillouin zone center with a capital letter $\vec{K} = (K_x, K_y,...)$.  
In contrast to the bosonic case, low energy modes live near an extended
surface (the Fermi Surface) rather than a single point (the Brillouin Zone center).
For a spherical Fermi surface, a high energy cutoff can be implemented 
on $\vec{K}$-integrals as follows:
\begin{eqnarray}
	\int^{\Lambda} d^dK
	&\equiv&
	\int d^{d-1}\Omega_{\vec{K}} \int_{K_F - \Lambda}^{K_F+\Lambda} K^{d-1}dK \nonumber
\end{eqnarray}
where $\Lambda$ is an ultraviolet cutoff, but we still insist $\Lambda \ll K_F$.
Here, $d^{d-1}\Omega_{\vec{K}}$ represents the measure for integration over all angular
coordinates in $\vec{K}$-space, while $K \equiv |\vec{K}|$ is the radial coordinate. 
Usually, we work at fixed fermion density which, by Luttinger's theorem, dictates 
that we design our scaling scheme in such a way that the Fermi volume remains invariant.  
To preserve the Fermi surface
under rescaling we cannot simply scale the radial coordinate as we did in the bosonic case.  To see this, observe that after mode elimination the expression we wish to rescale is given by 
\begin{eqnarray}
	\int^{\Lambda/s} d^dK
	&\equiv&
	\int d^{d-1}\Omega_{\vec{K}} \int_{K_F - \Lambda/s}^{K_F+\Lambda/s}K^{d-1}dK
\end{eqnarray}
Clearly, no simple rescaling of $K$ will return the integral to its original form.  This is the principle disparity between the fermionic and bosonic RG.  To make progress we define the lower case letter $k \equiv |\vec{K}|-K_F$.  Note that $k=0$ corresponds to $\xi_{\vec{K}}=\epsilon_{\vec{K}}-\mu=0$ since $\xi_{\vec{K}} = \frac{K^2-K_F^2}{2m} \approx v_F (K-K_F) = v_F k$.  Small $k$ corresponds to low energy whereas small $K$ does not.  Such a change of variables greatly facilitates rescaling.
\begin{eqnarray}
&&	\int d^{d-1}\Omega_{\vec{K}} \int_{- \Lambda/s}^{\Lambda/s}(K_F+k)^{d-1}dk  \nonumber\\
	&=&
	K_F^{d-1}\int d^{d-1}\Omega_{\vec{K}}\int_{- \Lambda/s}^{\Lambda/s}\left(1+\frac{k}{K_F}\right)^{d-1}dk  \nonumber \\
	&\approx&
	K_F^{d-1}\int d^{d-1}\Omega_{\vec{K}} \int_{- \Lambda/s}^{\Lambda/s}dk
\end{eqnarray}
We have neglected certain terms above for two reasons: they are of order $\Lambda/K_F$ 
relative to what has been kept, and they are less relevant in the RG sense.  To see the latter,
note that the integral can be restored to its original form with the simple rescaling 
$k^{\prime} =  sk$.  This determines the scaling dimension
\begin{eqnarray}
	\left[k\right] &=& 1
\label{eq:Scalingk}
\end{eqnarray}
Note that the variable $k$ is not a vector, nor is it a radial coordinate since it can take
negative values.  Later, we will discuss another scheme, which we call ``patching,'' that
decomposes the momenta into components parallel ($\vec{k}_{\parallel}$) and perpendicular
($k_{\perp}$) to the Fermi surface normal. 
To make later contrast with the patching scheme of section~\ref{sec:patching},
which uses local coordinates for each patch, we will call the present approach 
the ``global coordinate'' scheme.

To further emphasize the dissimilarity between the fermionic and bosonic cases, observe that after 
the rescaling of equation (\ref{eq:Scalingk}),
\begin{eqnarray}
	\int^{\Lambda/s} d^dK
	&\approx&
	K_F^{d-1}\int d^{d-1}\Omega_{\vec{K}} \int_{- \Lambda}^{\Lambda} s^{-1}dk^{\prime} \nonumber \\
	&=&
	s^{-1} \int^{\Lambda} (d^dK)^{\prime} 
\end{eqnarray}
which implies that, effectively,
\begin{eqnarray}
	\left[ d^dK \right] &=& 1
\end{eqnarray}
This stands in sharp contrast to the bosonic case in equation (\ref{eq:BosonMeasure}).
Here, the angular variables are truly untouched after rescaling which is necessary to maintain the Fermi surface.  
Unfortunately, the straightforward transformation $k^{\prime}=sk$ does not translate into a simple 
transformation on the components of $\vec{K}$.  Care must therefore be exercised to 
write all expressions in terms of $k$ before the scaling procedure can begin.  
For example, after mode elimination and rescaling of energy and momentum, 
the quadratic part of the fermionic action is given by:
\begin{eqnarray}
	\mathcal{S}_2^f 
	&\propto& s^{-3} \int
	dk^{\prime} d\epsilon^{\prime}
	\bar{\psi}(K_F + s^{-1}k^{\prime},s^{-1}i\epsilon^{\prime}) \nonumber\\
&&\times	
	\Big[ s^{-[\epsilon]} i\epsilon^{\prime} - v_F s^{-[k] } k^{\prime}\Big]
	\psi(K_F + s^{-1}k^{\prime},s^{-1}i\epsilon^{\prime}) \nonumber
\end{eqnarray}
If we wish to scale both of the terms inside the square brackets identically,
we must choose
\begin{eqnarray}
	\left[ k \right] &=& \left[ \epsilon \right]
\label{eq:keSame}
\end{eqnarray}
thus fixing the relationship between the scaling
dimensions of energy and momentum.
For convenience we can set this value equal to 1,
as in equation (\ref{eq:Scalingk}).
Compare this to equation (\ref{eq:Bosonwqz}).

In order to make $\mathcal{S}_2^f$ invariant to the RG transformation
we must demand that the fermion field obeys:
\begin{eqnarray}
\label{eq:FermionScaling}
	s^{-3/2}\psi(K_F + s^{-1}k^{\prime},s^{-1}i\epsilon^{\prime})
	&=&
	\psi^{\prime}(K_F + k^{\prime}, i\epsilon^{\prime})
\end{eqnarray}
where we have not explicitly written the dependence of $\psi$ on
angular variables since these do not scale.
Equation (\ref{eq:FermionScaling}) tells us two important things.
First, the dimension of the fermion field is simply:
\begin{eqnarray}
	\left[ \psi \right] &=& -3/2
\end{eqnarray}
Second, the RG transformation of the fermion field does {\it not} take
the form of a generalized homogeneous function as was the case
for the bosonic field; see equation (\ref{eq:BosonScaling}).
The momentum argument of the fermion field $\vec{K}$ has a magnitude equal to the 
Fermi wavevector plus a small deviation: $K = K_F +k$.  Only the deviation $k$ scales, while $K_F$ remains constant.  
This important difference from the bosonic case will be discussed further in section~\ref{sec:choiceOfScheme}.

The story so far seems relatively elementary, but the true subtleties materialize when we try to determine the dimension of the $\psi^4$ coupling function $u_f$ based on the dimension assignments required to make $\mathcal{S}_2^f$ scale invariant.  The quartic part of the action can be written~\cite{Shankar1994}
\begin{eqnarray}
	\mathcal{S}_4^f &=& \prod_{i=1}^4\int^{\Lambda} d^dK_i\int d\epsilon_i 
	\bar{\delta}^{(d)}(\vec{K}_1+\vec{K}_2 - \vec{K}_3 - \vec{K}_4) \nonumber\\
&&\times	\delta(\epsilon_1+\epsilon_2-\epsilon_3-\epsilon_4) \nonumber\\
&&\times	\bar{\psi}(4)\bar{\psi}(3)\psi(2)\psi(1) \;
	u_f(4,3,2,1)
\end{eqnarray}
The $\delta$-functions explicitly enforce the conservation of energy and momentum (up to a reciprocal lattice vector).  
We might integrate one of the energies and momenta, say $(\vec{K}_4,\epsilon_4)$, 
against the delta function to yield an integral over three independent sets 
$(\vec{K}_1,\epsilon_1)$, $(\vec{K}_2,\epsilon_2)$, and $(\vec{K}_3,\epsilon_3)$.
\begin{eqnarray}
	\mathcal{S}_4^f &=& \prod_{i=1}^3\int^{\Lambda} d^dK_i\int d\epsilon_i  
	\bar{\psi}(1+2-3)\bar{\psi}(3)\psi(2)\psi(1) \nonumber \\
	&&\times u_f(1+2-3,3,2,1) \;\; \text{ (wrong) }
\end{eqnarray}
But this expression is not quite right.  The problem is that not all momentum-conserving processes should be included in the low-energy effective field theory.  We must respect the cutoff imposed on the quadratic part of the action, which only allows excursion into states within a distance $\pm \Lambda$ of the Fermi surface.  Imposing a cutoff amounts to constraining the momentum integrals.  Until now, we have implemented the cutoff constraints by writing them explicitly in the limits of integration, but let us re-express them as 
\begin{equation}
\int^{\Lambda} d^dK_i = \int d^dK_i \Theta(\Lambda - |k_i|)
\end{equation}
where, as usual, $k_i \equiv |\vec{K}_i| - K_F$.  With all momentum integrals written in this way, we can safely use the $\delta$-functions to eliminate one variable, say $\vec{K}_4$ and $\epsilon_4$.
\begin{eqnarray}
	\mathcal{S}_4^f 
	&=& \prod_{i=1}^3\int d^dK_i\int d\epsilon_i  
	\bar{\psi}(1+2-3)\bar{\psi}(3)\psi(2)\psi(1) \nonumber \\
&&\times	 u_f(1+2-3,3,2,1)\nonumber\\
&&\times
	\Theta(\Lambda - |k_1|)
	\Theta(\Lambda - |k_2|)
	\Theta(\Lambda - |k_3|)
	\Theta(\Lambda - |\mathcal{K}_4|) \nonumber \\
	&=& \prod_{i=1}^3\int^{\Lambda} d^dK_i\int d\epsilon_i  
	\bar{\psi}(1+2-3)\bar{\psi}(3)\psi(2)\psi(1) \nonumber \\
	&&\times	 u_f(1+2-3,3,2,1) \Theta(\Lambda - |\mathcal{K}_4|)
	\label{eq:fourFermion}
\end{eqnarray}
The constraints on $\vec{K}_1$, $\vec{K}_2$, and $\vec{K}_3$ have been put back in the limits of integration, but we have the additional constraint $|\mathcal{K}_4| < \Lambda$ where 
\begin{equation}
	\mathcal{K}_4 \equiv |\vec{K}_3-\vec{K}_2-\vec{K}_1| - K_F
	\label{eq:KFOUR}
\end{equation}
Once we have conserved momentum, $\vec{K}_4$ is no longer an independent variable, so we use the notation
$\mathcal{K}_4$ to represent the combination of variables specified in equation (\ref{eq:KFOUR}).

We can implement the constraint embodied in $\Theta(\Lambda - |\mathcal{K}_4|)$ in a number of ways.  
One way is to allow $\vec{K}_1$ and $\vec{K}_2$ to range anywhere inside the annuli defined by $-\Lambda < k_1,k_2 < \Lambda$, but restrict $\vec{K}_3$ as appropriate to satisfy $|\mathcal{K}_4| < \Lambda$.  The outcome of a proper phase space analysis shows that once $\vec{K}_1$ and $\vec{K}_2$ have been chosen, the angle for $\vec{K}_3$ is highly constrained~\cite{Shankar1994}.

To see this in more detail, observe that to leading order in $\Lambda/K_F$, 
\begin{eqnarray}
	\mathcal{K}_4 &\approx& K_F(|\vec{\Delta}|-1)
	\label{eq:KFOURapprox}
\end{eqnarray} 
where 
$\vec{\Delta} \equiv \hat{K}_1+\hat{K}_2-\hat{K}_3$, 
and where the $\hat{K}_i$ are unit vectors, each pointing in the direction of $\vec{K}_i$.
Note that $\vec{\Delta}$ is not itself a unit vector since
\begin{eqnarray}
	|\vec{\Delta}|
	&=& 
	\sqrt{2}\left[\frac{3}{2}+\hat{K}_1\cdot\hat{K}_2 - \hat{K}_1\cdot\hat{K}_3 - \hat{K}_2\cdot\hat{K}_3 \right]^{1/2}
\label{eq:modDelta}
\end{eqnarray}
a result we will use in section~\ref{sec:choiceOfScheme}.
After mode elimination, the momentum integrals become:
\begin{eqnarray}
	&&\prod_{i=1}^3\int d^dK_i \Theta\left(\Lambda/s - K_F\left| |\vec{\Delta}|-1\right|\right) \nonumber \\
	&&=
	\prod_{i=1}^3\int d^dK_i \Theta\left(\Lambda - s K_F\left| |\vec{\Delta}|-1\right| \right)
\label{eq:fermionConstraint}
\end{eqnarray}
Simply rescaling $k^{\prime}_i = sk_i$ is not sufficient to regain the original form of the action for generic values of $k_i$.  The obvious snag is the annoying way the $\Theta$-function transforms.  
For general values of the momenta $k_i$, the $\Theta$-function is clearly not invariant to the renormalization group transformation.  
Consequently, we are not technically entitled to compare the coupling before and after, so we do not know the RG flow.
The way out of this dilemma is first to understand the circumstances under which the $\Theta$-function
is invariant, and then to see what might be happening for more generic cases by considering a soft cutoff.

First, note that when $|\vec{\Delta}| = 1$ (i.e. $\mathcal{K}_4=0$) the $\Theta$-function is always form-invariant
since $\Theta(\Lambda) = \Theta(\Lambda/s)$.  
The condition $|\vec{\Delta}|=1$ can be fulfilled in three different ways:
\begin{eqnarray}
	\label{eq:forward}	&\text{(i)}&   \vec{K}_1=\vec{K}_3 \text{ and } \vec{K}_2=\vec{K}_4 \\
	\label{eq:exchange}	&\text{(ii)}&  \vec{K}_2=\vec{K}_3 \text{ and } \vec{K}_1=\vec{K}_4 \\
	\label{eq:cooper}	&\text{(iii)}& \vec{K}_1=-\vec{K}_2 \text{ and } \vec{K}_3=-\vec{K}_4
\end{eqnarray}
For these values of the momenta, the rescaling $k^{\prime}_i = sk_i$ works flawlessly 
because the $\Theta$-function is form-invariant under these restrictions.
We are now allowed to compare the coupling before and after.
Since 
$[ dk_1 dk_2 dk_3 d\epsilon_1d\epsilon_2d\epsilon_3] = 6$ and
$[ \psi^4] = -6$, we conclude that, at the tree level, the most relevant pieces of $u_f$ are marginal.
This important result is at the heart of Fermi Liquid Theory, but is expressed by the simple equation:
\begin{eqnarray}
	\left[ u_f \right] &=& 0
\end{eqnarray}
In Shankar's notation, cases (i) and (ii) correspond to $u_f=\pm F$ and case (iii) $u_f=V$.
It has also been shown~\cite{Shankar1994} that case (i) remains marginal beyond the tree level, while loop
corrections in case (iii) lead to a marginally relevant coupling for certain angular momentum channels, 
indicative of the BCS instability.

\begin{figure}[htbp]
   \centering
   \includegraphics[width=3in]{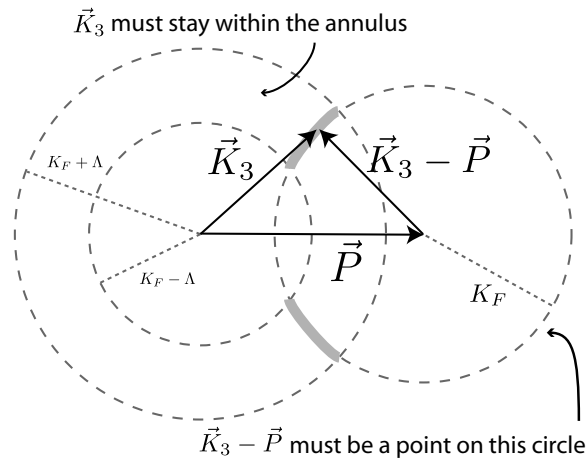}
   \caption[Kinematic constraints of the four-fermion coupling.]
   {Once $\vec{K}_1$ and $\vec{K}_2$ have been chosen
   the conservation of momentum and the requirement that all $\vec{K}_i$
   respect the cutoff of the field theory strongly constrains the phase space
   available to $\vec{K}_3$.  Shown here is the limit $\mathcal{K}_4 =0$.
   Even in this limit, there is some flexibility in the choices available to $\vec{K}_3$, 
   as depicted by the thick gray lines.  While the magnitude of $k_3$ can still
   fall anywhere in the range $-\Lambda < k_3 < \Lambda$, the angles available
   to $\vec{K}_3$ are highly limited.  If we were to further take the limit $\Lambda/K_F \to 0$,
   the gray lines would shrink to points.
   Note that $\vec{P}$ is defined as the sum of $\vec{K}_1$ and $\vec{K}_2$,
   but the latter are not drawn to avoid clutter.}
   \label{fig:shankarPhaseSpace}
\end{figure}

Let us understand in more detail the circumstances under which the $\Theta$-function is
always form-invariant.  In particular, we want to stress that the
the condition $\mathcal{K}_4 = 0$ 
is conceptually different from the limit $\Lambda/K_F \to 0$.  
To see this, let us rewrite the equation $\mathcal{K}_4=0$ as follows:
\begin{eqnarray}
	|\vec{K}_3 - \vec{K}_2 - \vec{K}_1 | &=& K_F
\end{eqnarray}
Note that equation (\ref{eq:KFOUR}) is slightly more accurate than (\ref{eq:KFOURapprox}).
Next, define $\vec{P} \equiv \vec{K}_1 + \vec{K}_2$ which obviously gives
\begin{eqnarray}
	|\vec{K}_3 - \vec{P}| &=& K_F
\end{eqnarray}
This says that the vector joining the tip of $\vec{K}_3$ to the tip of $\vec{P}$ must have magnitude precisely equal to $K_F$.  
Figure~\ref{fig:shankarPhaseSpace} depicts the situation.
Geometrically, the choices available to $\vec{K}_3$ once $\vec{K}_2$ and $\vec{K}_1$ have been selected are given by the thick gray lines in the figure.  Notice that while $k_3$ can still take any values $-\Lambda < k_3 < \Lambda$ within the annulus, the angle of $\vec{K}_3$ has become highly constrained.  
However, it is clear that even when $\mathcal{K}_4 = 0$ the value of $\Lambda/K_F$ can still be nonzero.

We now know the dimension $[u_f]$ when we restrict to $\mathcal{K}_4 = 0$.
However, the three cases corresponding to $\mathcal{K}_4 = 0$ constitute only a small portion of $(\vec{K}_1, \vec{K}_2, \vec{K}_3)$-space.  To see what happens to the coupling function $u(3,2,1)$ for other values of momenta, Shankar had the insight to employ a soft cutoff: $\Theta(\Lambda - |k_i|) \approx e^{-|k_i| / \Lambda} $.  Using this device, the rescaled cutoff for arbitrary $k_i$ values becomes
\begin{eqnarray}
	\Theta\left(\Lambda - s K_F\left| |\vec{\Delta}|-1\right| \right) 
		&\approx& e^{-sN_{\Lambda}\left| |\vec{\Delta}| - 1 \right| } \nonumber \\
		&=& e^{-N_{\Lambda}\left| |\vec{\Delta}| - 1 \right| } e^{-(s-1)N_{\Lambda}\left| |\vec{\Delta}| - 1 \right| } \nonumber
\end{eqnarray}
where we have defined the large parameter $N_{\Lambda} \equiv K_F/\Lambda$ (generally, we have the hierarchy $k < \Lambda \ll K_F$, which means $N_{\Lambda} \gg 1$).  We choose to write the cutoff in this way because then clearly when $|\vec{\Delta}|=1$, corresponding to the three cases listed above, the cutoff becomes a simple factor of unity.  For any $|\vec{\Delta}| \neq 1$, which means all other values of the $k_i$, the cutoff $\to 0$ in the limit $N_{\Lambda} \to \infty$ provided $s>1$.  While we do not know how the coupling function $u(3,2,1)$ scales for values of the momenta where $|\vec{\Delta}| \neq 1$, it does not matter because these couplings will be exponentially suppressed in the limit $1/N_{\Lambda} = \Lambda/K_F \to 0$.

Note that the condition $|\vec{\Delta}| = 1$ is simply the statement 
that $\mathcal{K}_4$ should not scale.  Indeed, it means $\mathcal{K}_4 = 0$.
Only when $|\vec{\Delta}| = 1$ is the relation $\mathcal{K}_4^{\prime} = 
s \mathcal{K}_4$ satisfied, albeit trivially.  
In fact, that is how we identified the condition $|\vec{\Delta}| = 1$,
being the only combination of
$\vec{K}_3$, $\vec{K}_2$, and $\vec{K}_1$ 
where $\Theta(\Lambda - |\mathcal{K}_4|)$
can be rescaled to take its original form after mode elimination.
This useful interpretation will be used again later when we extend
the formalism to include bosons.

Before moving on, we need to make another observation about the pure fermion RG
that will be important to later generalizations.  We have shown how to find the dimension
of the coupling function $u_f(3,2,1)$ for those values of momentum that satisfy
$\mathcal{K}_4 = 0$ (\textit{i.e.} $|\hat{\Delta}| = 1$) corresponding to 
forward, exchange, and Cooper scattering.  To be pedantic, this phase space
restriction should be incorporated into the form of the coupling:
\begin{eqnarray}
\label{eq:S4Constrained}
	\mathcal{S}_4^f 
	&=& \prod_{i=1}^3\int^{\Lambda} d^dK_i\int d\epsilon_i  
	\bar{\psi}(1+2-3)\bar{\psi}(3)\psi(2)\psi(1) \nonumber \\
&&\times	 
	u_f(1+2-3,3,2,1) 
	\Theta(\Lambda - |\mathcal{K}_4|)
	\delta(\mathcal{K}_4)
\end{eqnarray}
Note that $\Theta(\Lambda) = 1$ always since $\Lambda >0$.
As seen in figure~\ref{fig:shankarPhaseSpace}, the insertion of
$\delta(\mathcal{K}_4)$  
does not affect the freedom of
$\vec{K}_1$ or $\vec{K}_2$ at all, nor does it affect the magnitude of $\vec{K}_3$
so long as $-\Lambda < |\vec{K}_3| - K_F < \Lambda$.  However, the angle
of $\vec{K}_3$ is highly restricted to the two gray regions of the figure as a resulting
of inserting $\delta(\mathcal{K}_4)$.  
We may therefore implement the constraint (in $d=2$) by:
\begin{eqnarray}
\label{eq:theta3Constraint}
	\delta(\mathcal{K}_4) 
	&\to& \delta (|\vec{\Delta}|-1)/K_F \nonumber\\
	&&=[\delta(\theta_3 - \theta_1) + \delta(\theta_3 -\theta_2)]/K_F 
\end{eqnarray}
A similar expression can be written in $d=3$.
Since angles do not scale in this scheme, whether or not we insert
this factor into $\mathcal{S}_4^f$ will have \textit{no effect} on the
value of the dimension of $u_f$.
Shankar's result of marginality, $[u_f] = 0$, still holds.
We mention this issue because
generalizing the method to include bosons will not result
in so happy a circumstance.  We turn to this case next.

\section{Boson+Fermion Scaling}
\label{sec:bosons+fermions}

We are finally ready to incorporate bosons.  Consider the following interaction term involving two fermions and one boson:
\begin{eqnarray}
	\mathcal{S}_3^{bf} 
	&=& \int d^dK_1 d^dK_2 d^dq \; g(\vec{K}_1,\vec{K_2},\vec{q}) \; \bar{\psi}_{\vec{K}_2}\psi_{\vec{K}_1} \phi_{\vec{q}} \nonumber \\
	&&\times \delta^{(d)}( \vec{K}_2  - \vec{K}_1 - \vec{q} )
	\Theta(\Lambda-|k_1|)\Theta(\Lambda-|k_2|) \nonumber \\
	&&\times \Theta(\Lambda-|\vec{q}|)
\label{eq:BFCoupling}
\end{eqnarray}
$g$ is the coupling function which plays the same role as $u_f$ in the 4-fermion problem.  For simplicity we have suppressed frequency integrals and assumed $\Lambda_b \sim \Lambda_f \sim \Lambda$.  
To conserve momentum we have two choices: use the $\delta$-function to eliminate a fermionic momentum $\vec{K}_i$, or the bosonic momentum $\vec{q}$. This gives either
\begin{eqnarray}
	\int^{\Lambda} d^dK d^dq\left[ \bar{\psi}_{\vec{K}+\vec{q}}\psi_{\vec{K}} \phi_{\vec{q}}\,g(\vec{K},\vec{q})
	\Theta(\Lambda - |\mathcal{K}_2|)\right]
\label{eq:KQScheme}
\end{eqnarray}
or
\begin{eqnarray}
	\int^{\Lambda} d^dK_1 d^dK_2 \left[\bar{\psi}_{\vec{K}_2}\psi_{\vec{K}_1} \phi_{K_2-K_1}\,g(\vec{K}_2,\vec{K}_1)
	\Theta(\Lambda - |\vec{\mathcal{Q}}|)\right] \nonumber \\
\label{eq:K1K2Scheme}
\end{eqnarray}
where some of the cutoff constraints have been put back in the limits of integrations, 
and where we have defined
\begin{eqnarray}
	\mathcal{K}_2 &\equiv& |\vec{K}+\vec{q}| - K_F \label{eq:K2}\\
	\vec{\mathcal{Q}} &\equiv& \vec{K}_1 - \vec{K}_2 \label{eq:Q}
\end{eqnarray}
This is analogous to what we did for the pure fermion problem; 
see equations (\ref{eq:fourFermion}) and (\ref{eq:KFOUR}).  
Note that because we integrated against the delta functions, momentum and energy are already explicitly conserved.  
In equation (\ref{eq:KQScheme}), $\vec{K}_2$ is no longer an independent variable, so we use the symbol
$\mathcal{K}_2$ to represent the combination of variables specified in equation (\ref{eq:K2}).
Likewise, $\vec{q}$ is not an independent variable in equation (\ref{eq:K1K2Scheme}),
so we use $\vec{ \mathcal{Q} }$ as shorthand for the momentum transfer, as specified in equation $(\ref{eq:Q})$.
This mirrors the development of the pure fermion case.

Unlike the pure fermion problem,
we now appear to have two different choices for expressing the boson-fermion
coupling.  Equation (\ref{eq:KQScheme}) involves the boson-fermion coupling function
$g(\vec{K},\vec{q})$, while equation (\ref{eq:K1K2Scheme}) contains $g(\vec{K}_2,\vec{K}_1)$.
We defer a discussion of the resolution of this choice to section~\ref{sec:choiceOfScheme}.
Here, we simply point out that a consistent scheme can only be found 
for equation (\ref{eq:KQScheme}),
and we adopt this choice for the remainder of this section.

Although momentum is conserved, just like the pure fermion case, not all momentum conserving processes are allowed because some might fall outside the high-energy cutoffs.  We must further restrict the coupling function $g$ with the constraint $\Theta(\Lambda - |\mathcal{K}_2| )$.  Unfortunately, this quantity only scales in a simple way when $z=1$.  Let us briefly explain the problem.

Recall from the form of $\mathcal{S}_2^f$ that we have the relation $[k]=[\epsilon]$, while $\mathcal{S}_2^b$ demands $[q]=[\omega]/z$ for general values of $z$ 
[see equations (\ref{eq:keSame}) and (\ref{eq:Bosonwqz})].  
In addition, since we want to scale fermions and bosons at the same time, we choose to scale the energies the same way, that is: $[\omega] = [\epsilon]$.  For convenience, we set the scaling dimension of energy to unity: $[\omega] = [\epsilon] = 1$.  Any other value would change all scaling dimensions by the same multiplicative factor, but their relative dimensions would be unaffected.  
Using this prescription we find
\begin{eqnarray}
\label{eq:enDim}		
\left[\epsilon\right] &=& \left[k\right] = 	\left[\omega\right] = 1
\nonumber 
\\
					\left[q\right] &=& \left[\omega\right]/z = \frac{1}{z} 
\nonumber \\
\label{eq:fermionDim}	\left[\psi\right] &=& - \frac{3}{2} 
\nonumber 
\\
\label{eq:bosonDim}		\left[\phi\right] &=& -\frac{d+z+2}{2z}
\end{eqnarray}
Mode elimination and rescaling according to this scheme leads to the following interaction term (we reinstate the energy integrals):
\begin{eqnarray}
	\mathcal{S}_3^{bf} 
	&=& s^{\frac{z + 2 - d}{2z} } 
	g \int^{\Lambda} d^dq^{\prime} dk^{\prime} d^{d-1}\Omega_{\vec{K}} 
	d\epsilon^{\prime}d\omega^{\prime} 
	\bar{\psi}^{\prime}\psi^{\prime}\phi^{\prime} \nonumber \\ 
	&&\times \Theta\Big( \Lambda/s - |\mathcal{K}_2 | \Big)
\label{eq:S3}
\end{eqnarray}
The reason why we have $\Lambda/s$ rather than $\Lambda/s^{1/z}$
is because this constraint comes from the restriction on the 
momentum integration of $k_2 \equiv |\vec{K}_2| - K_F$ in 
equation (\ref{eq:BFCoupling}), which scales like a fermion.

Let us rewrite the expression involved in the $\Theta$-function:
\begin{eqnarray}
	| \mathcal{K}_2 |
		&=& |\vec{K}+\vec{q}| - K_F \nonumber \\
		&=& \left[ (K_F+k)^2 + q^2 + 2(K_F+k)q\cos\theta_{Kq}\right]^{1/2} - K_F \nonumber \\
		&\approx& K_F\left[ 1+\frac{2}{K_F}(k+q\cos\theta_{Kq}) \right]^{1/2} - K_F \nonumber \\
		&\approx& k+q\cos\theta_{Kq} 
\end{eqnarray}
which is valid to leading order in $\Lambda/K_F$, and where
\begin{eqnarray}
	\cos\theta_{Kq} 
	&=& \hat{K}\cdot \hat{q} \\
	&\equiv& \left\{
	\begin{array}{l}
		\cos(\theta_K - \theta_q) \\ 
		\cos\theta_K\cos\theta_q + \sin\theta_K\sin\theta_q\cos(\varphi_K - \varphi_q) 
	\end{array}
	\right. \nonumber\\
\end{eqnarray}
in $d=2$ and $d=3$, respectively.
Equation (\ref{eq:S3}) now becomes
\begin{eqnarray}
	\mathcal{S}_3^{bf} 
	&=& s^{\frac{z + 2 - d}{2z} } 
	g \int^{\Lambda} d^dq^{\prime} dk^{\prime} d^{d-1}\Omega_{\vec{K}} 
	d\epsilon^{\prime}d\omega^{\prime} 
	\bar{\psi}^{\prime}\psi^{\prime}\phi^{\prime} \nonumber \\ 
	&&\times \Theta\Big( \Lambda/s - | k+q\cos\theta_{Kq} | \Big) \nonumber \\
	&=& s^{\frac{z + 2 - d}{2z} } 
	g \int^{\Lambda} d^dq^{\prime} dk^{\prime} d^{d-1}\Omega_{\vec{K}} 
	d\epsilon^{\prime}d\omega^{\prime} 
	\bar{\psi}^{\prime}\psi^{\prime}\phi^{\prime} \nonumber \\ 
	&&\times \Theta\Big( \Lambda - | k^{\prime}+s^{(z-1)/z}q^{\prime}\cos\theta_{Kq} | \Big)
	\label{eq:S3zneqOne}
\end{eqnarray}
where 
$k^{\prime} = s k$, 
$q^{\prime} = s^{1/z}q$, 
$\epsilon^{\prime} = s\epsilon$, and 
$\omega^{\prime} = s\omega$.
Clearly, for generic values of $z$ 
the $\Theta$-function does not return to its original form 
after the renormalization group transformation.
We should be pleased, however, that in the special case $z=1$, 
the $\Theta$-function is form-invariant.
\begin{eqnarray}
	\Theta(\Lambda/s - |\vec{\mathcal{K}_2}|) 
		&\approx& \Theta(\Lambda/s - | k+q\cos\theta_{Kq} | ) \nonumber \\
		&=& \Theta(\Lambda - s| k+q\cos\theta_{Kq} | ) \nonumber \\
		&=& \Theta(\Lambda - | k^{\prime}+q^{\prime}\cos\theta_{Kq} | ) 
\end{eqnarray}
The boson-fermion coupling can now be written
\begin{eqnarray}
	\mathcal{S}_3^{bf} 
	&=& s^{\frac{3-d}{2} } 
	g \int^{\Lambda} d^dq^{\prime} dk^{\prime} d^{d-1}\Omega_{\vec{K}} 
	d\epsilon^{\prime}d\omega^{\prime} 
	\bar{\psi}^{\prime}\psi^{\prime}\phi^{\prime} \nonumber \\
&&\times	\Theta\Bigg( \Lambda - |k^{\prime}+q^{\prime}\cos\theta_{Kq}| \Bigg)
\end{eqnarray}
and we can identify
\begin{eqnarray}
	g^{\prime} \equiv s^{(3-d)/2} g
\end{eqnarray}
which is equivalent to 
\begin{eqnarray}
	\left[ g \right] &=& (3-d)/2
\label{eq:gDimzEqualsOne}
\end{eqnarray}
This is one of the central results of this paper.
The coupling is marginal in $d=3$ and relevant in $d=2$.   
Of course, this result depends on the choice of field dimensions;
equations (\ref{eq:bosonDim}) with $z=1$.
In application to an antiferromagnetic Kondo lattice, we have previously
developed a model where the boson dimension is $-d$
[rather than equation (\ref{eq:bosonDim})]
and used the scheme explained here to show that the boson-fermion coupling is
exactly marginal in that case~\cite{Yamamoto2007}.

Equation (\ref{eq:gDimzEqualsOne}) is only valid when $z=1$ because only
then is the $\Theta$-function form-invariant.  What can be done when $z \neq 1$?
This question is particularly pertinent to the controversy surrounding the
renormalization of a gauge-field coupled to a fermion with a Fermi surface
~\cite{Polchinski1994, Nayak1994, Altshuler1994, Onoda1995}.
It is also germane to
the ferromagnetic phase of heavy fermion systems~\cite{Yamamoto2008b}.
Section~\ref{sec:patching} will present a different scheme that is applicable
to problems with arbitrary values of $z$ and also reproduces 
equation (\ref{eq:gDimzEqualsOne}) when $z=1$.
Here, we merely explain why the present scheme fails when $z \neq 1$.

\begin{figure}[htbp]
   \centering
   \includegraphics[width=3in]{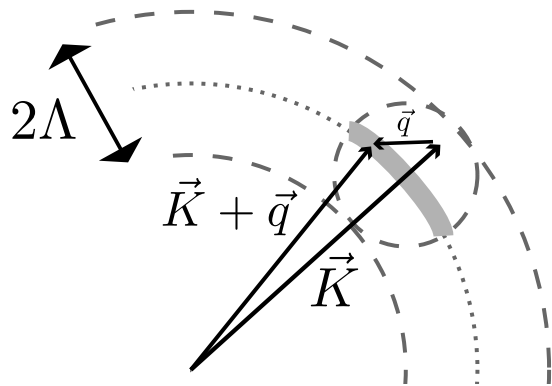}
   \caption[Kinematic constraints of the boson-fermion coupling]
   {$\vec{K}$ must stay within the annulus while $\vec{q}$ must stay inside the little circle of radius $\Lambda$.  Under the restriction $\mathcal{K}_2 = 0$, the sum $\vec{K}+\vec{q}$ must sit precisely on the Fermi surface.  The only phase space that satisfies $\mathcal{K}_2 = 0$ is the thick gray line which represents a small patch on the Fermi surface of size O($\Lambda^{d-1}$).  Clearly, the limit $\mathcal{K}_2=0$ is not the same as $\Lambda/K_F =0$ since the latter would shrink the gray patch to a point.  }
   \label{fig:KQPhaseSpace}
\end{figure}

We have actually already seen the problem in equation (\ref{eq:S3zneqOne}),
where it is obvious that the $\Theta$-function is not form invariant.
This is similar to the dilemma we encountered in the pure
fermion problem, as seen in equation (\ref{eq:fermionConstraint}).
To make progress, we try the same strategy used in the 
pure-fermion problem where we restricted our consideration to the phase space 
where the $\Theta$-function does scale perfectly.
We did so by demanding $\mathcal{K}_4=0$ (or $|\vec{\Delta}|=1$) in 
$\mathcal{S}_4^f$, which can be implemented by simply inserting $\delta(\mathcal{K}_4)$.
Here, the analogue of that additional constraint is $\mathcal{K}_2 = 0$.  
This new condition can also be written
\begin{eqnarray}
	|\vec{K}+\vec{q}| &=& K_F
\end{eqnarray}
Thus, besides staying within their respective cutoffs, 
the choices available to $\vec{K}$ and $\vec{q}$ ,
when $\mathcal{K}_2 = 0$, are restricted in such a way that their sum 
vector must sit precisely on the Fermi surface.  
Once $\vec{K}$ is chosen, $\vec{q}$ is obligated to connect 
$\vec{K}+\vec{q}$ to the Fermi surface thus limiting its permissible 
magnitudes and angles quite severely.
This is depicted in figure~\ref{fig:KQPhaseSpace}.

Under the restriction $\mathcal{K}_2 = 0$, 
the boson-fermion coupling can be written:
\begin{eqnarray}
	\mathcal{S}_3^{bf} 
	&=& s^{\frac{z + 2 - d}{2z} } 
	g \int^{\Lambda} d^dq^{\prime} dk^{\prime} d^{d-1}\Omega_{\vec{K}} 
	d\epsilon^{\prime}d\omega^{\prime} 
	\bar{\psi}^{\prime}\psi^{\prime}\phi^{\prime} \nonumber \\ 
	&&\times \Theta\big( \Lambda/s - |\mathcal{K}_2| \big)
	\delta\big( \mathcal{K}_2 \big) \\
	&=& s^{\frac{z + 2 - d}{2z} } 
	g \int^{\Lambda} d^dq^{\prime} dk^{\prime} d^{d-1}\Omega_{\vec{K}} 
	d\epsilon^{\prime}d\omega^{\prime} 
	\bar{\psi}^{\prime}\psi^{\prime}\phi^{\prime} \nonumber \\ 
	&&\times	\delta\Big( k^{\prime}+s^{(z-1)/z}q^{\prime}\cos\theta_{Kq}\Big) 
	\label{eq:S3K2EqualsZero}
\end{eqnarray}
This should be compared with equation (\ref{eq:S4Constrained}).
In the pure fermion case, we showed there that whether or not we
insert $\delta(\mathcal{K}_4)$ makes no difference to the value of
$[u_f]$ because $\delta(\mathcal{K}_4)$ is $\sim \delta(\theta_3 - \theta_1)$
or $\sim \delta(\theta_3 - \theta_2)$,
which has zero scaling dimension.  Furthermore, this additional constraint
is of a non-singular nature.

In contrast, for the boson-fermion
coupling in equation (\ref{eq:S3K2EqualsZero}), the insertion of
$\delta(\mathcal{K}_2)$ involves a dimensionful quantity.
If we were to integrate against $\delta(\mathcal{K}_2)$ and eliminate
$\theta_q$ as suggested by figure~\ref{fig:KQPhaseSpace}, we would
induce an additional 1/momentum factor in violation of the
RG edict that the coupling be a non-singular function of momentum.
More intuitively, figure~\ref{fig:KQPhaseSpace} shows that imposing the
constraint $\mathcal{K}_2 = 0$ singles out an unrealistic sort of coupling
that glues the out-going fermion $\bar{\psi}_{\vec{K}+\vec{q} }$ to the Fermi
surface regardless of the value of $\vec{K}$ or $\vec{q}$.   This no longer
represents a generic forward scattering process, and is of no interest to us.
How to correctly capture a generic forwarding scattering process will be
discussed in section~\ref{sec:patching}.

At this point, a few issues are worth emphasizing.
\begin{itemize}
	\item Since we integrated against the delta functions in equation (\ref{eq:BFCoupling}), 
	energy and momentum are explicitly conserved.

	\item The quantity $\mathcal{K}_2$ is not a free variable and it does not necessarily scale in the same way as bosonic or fermionic momenta.  This is consistent with the non-scaling of $\mathcal{K}_4$ in the pure-fermion problem when $|\vec{\Delta}| \neq 1$.  Only when $z=1$ does $\mathcal{K}_2$, and thus the constraint $\Theta(\Lambda - |\mathcal{K}_2|)$, scale in a simple way.

	\item In this scheme, all components of $\vec{q}$ scale the same way.  In particular, $[d^dq]=d/z=d$.  At the same time, only fermionic momenta in the direction normal to the Fermi surface scale.

	\item Here, $k$ is not a vector.  It does not have parallel or perpendicular 
components as discussed in certain patching schemes.  For more on the patching scheme,
see section~\ref{sec:patching}.

	\item Although the RG scheme developed in this section does not work for general values of $z$,
it is perfectly well suited to the special case $z=1$. 
 
	\item Figure~\ref{fig:KQPhaseSpace} gives us an important hint about what may be happening
for $z>1$.  Since $k^{\prime} = sk$ and $q^{\prime} = s^{1/z}q$, we know that after several iterations
of the RG, the deviation of $\vec{K}$ from the Fermi surface will be much smaller than the
magnitude of $\vec{q}$, {\it i.e.} $k \ll q$.  As a result $\vec{q}$
will tend to point in a direction perpendicular to $\vec{K}$, which means it will be
very nearly tangent to the Fermi surface.  In this way, it may seem as if bosonic momenta
scale anisotropically in a local coordinate system defined with respect to the direction
determined by a fixed $\vec{K}$.  This important observation will be developed more
fully in section~\ref{sec:patching} when we devise a scheme suitable to $z \neq 1$.
\end{itemize}

\section{Choice of Momentum Integration}
\label{sec:choiceOfScheme}

Before moving on to the general case $z \neq 1$,
in this section we resolve a seeming ambiguity for the
scheme we developed in the previous section.
As we found in equations 
(\ref{eq:KQScheme})
and
(\ref{eq:K1K2Scheme})
there are two ways to express $\mathcal{S}_3^{bf}$ in
momentum space.  We have already shown in detail
that making the choice in equation (\ref{eq:KQScheme})
can yield a consistent RG prescription.
Now we will show why the alternative decomposition
\begin{eqnarray}
	\int^{\Lambda} d^dK_1 d^dK_2 \left[\bar{\psi}_{\vec{K}_2}\psi_{\vec{K}_1} \phi_{K_2-K_1}\,g(\vec{K}_2,\vec{K}_1)
	\Theta(\Lambda - |\vec{\mathcal{Q}} |) \right] \nonumber
\end{eqnarray}
is not an appropriate starting point to determine the scaling dimension of the boson-fermion coupling.  
The problem is that the argument of the boson field, $\vec{\mathcal{Q}} \equiv \vec{K}_1-\vec{K}_2$, 
does not transform homogeneously, so we do not know what dimension to assign to the boson itself.
To see this, write each fermion momentum vector in terms of a direction
and a deviation from the Fermi surface: $\vec{K}_i = (K_F + k_i)\hat{K}_i$.
This gives
\begin{eqnarray}
	| \vec{ \mathcal{Q} } |
		&=& \left[ K_1^2 + K_2^2 - 2K_1K_2\cos\theta_{12} \right]^{1/2} \nonumber \\
		&=& \Big[ (K_F + k_1)^2 + (K_F + k_2)^2	- 2(K_F+k_1) \nonumber\\
		&&\times(K_F + k_2)\cos\theta_{12} \Big]^{1/2} \nonumber \\
		&\approx& K_F \sqrt{2}\left[(1-\cos\theta_{12})\left(1+\frac{k_1+k_2}{K_F}\right) \right]^{1/2} \nonumber\\
		&\approx& K_F \sqrt{2(1-\cos\theta_{12})} \left(1+\frac{k_1+k_2}{2K_F} \right)
\label{eq:approxQ}
\end{eqnarray}
which is true to leading order in $1/N_{\Lambda}$, and where
\begin{eqnarray}
	\cos\theta_{12} 
	&=& \hat{K}_1 \cdot \hat{K}_2 
\nonumber \\
	&\equiv& \left\{
	\begin{array}{ll}
		\cos(\theta_1 - \theta_2) \\
		\cos\theta_1\cos\theta_2+ \sin\theta_1\sin\theta_1\cos(\varphi_1 - \varphi_2) 
	\end{array}\right. \nonumber \\
\end{eqnarray}
for $d=2$ and $d=3$.
In equation (\ref{eq:FermionScaling})
we committed to a specific prescription in making $\mathcal{S}_2^f$ scale invariant 
where angular components of the momentum do not scale.
We therefore cannot allow angles to scale in $\mathcal{S}_3^{bf}$
either.  
Using the specific prescription 
in equations (\ref{eq:enDim}) determined by the
quadratic parts of the action we find
\begin{eqnarray}
	\mathcal{S}_3^{bf}
	&=&
	\int^{\Lambda} 
	s^{-1}dk_1^{\prime} s^{-1}dk_2^{\prime} 
	s^{-1}d\epsilon_1^{\prime} s^{-1}d\epsilon_2^{\prime} \nonumber\\
&&	\Bigg[
	s^{3/2}\bar{\psi}^{\prime}(K_F + k_2^{\prime}, i\epsilon_2^{\prime})
	s^{3/2}\psi^{\prime}(K_F + k_1^{\prime}, i\epsilon_1^{\prime}) \nonumber \\
	&&\times 
	\phi\Big(K_F \sqrt{2(1-\cos\theta_{12})} \left[1+\frac{k_1^{\prime}+k_2^{\prime}}{2sK_F} \right], \nonumber \\
&&	s^{-1}i\epsilon_2^{\prime}-is^{-1}\epsilon_1^{\prime}\Big)
	g(2,1)
	\Theta(\Lambda/s^{1/z} - |\vec{ \mathcal{Q} } |) \Bigg]
\label{eq:S3K1K2Rep}
\end{eqnarray}
where
\begin{eqnarray}
	\Theta(\Lambda/s^{1/z} - |\vec{\mathcal{Q}} |) 
	&\equiv&
	\Theta\Bigg(\Lambda - s^{1/z}K_F \sqrt{2(1-\cos\theta_{12})} \nonumber\\
	&&\times\left[1+\frac{k_1^{\prime}+k_2^{\prime}}{2sK_F} \right]\Bigg)
\label{eq:K1K2Constraint}
\end{eqnarray}
Notice that in equation (\ref{eq:S3K1K2Rep}) the fermion fields are primed whereas
the boson field is not.

There are two problems.  First, the $\Theta$-function does not return to its original form,
making it impossible to compare the flow of the coupling function before and after 
the RG transformation.  This is the same problem we encountered in the pure fermion
case of section~\ref{sec:fermions}, 
and the other boson-fermion prescription from section~\ref{sec:bosons+fermions}.

Second, we have a new dilemma, which is that we do not know how the $\phi$ field transforms under
the change of argument in equation (\ref{eq:S3K1K2Rep}).  
All we know from equation (\ref{eq:BosonScaling}) is that
\begin{eqnarray}
	\phi^{\prime}(q^{\prime},i\omega^{\prime}) \equiv s^{-(d+z+2)/(2z)} \phi(s^{1/z} q, s i\omega)
	\label{eq:phiScaling}
\end{eqnarray}
which states that the boson scales in a (generalized) homogeneous fashion.  
If we transform the boson
arguments in a non-homogeneous way, as in (\ref{eq:S3K1K2Rep}), 
we are not guaranteed that such a coordinate transformation
will induce a simple multiplicative prefactor. 
Note that the mathematical requirement that the boson field transform homogeneously means
that the relative angle between the incoming and outgoing fermions must be allowed to scale.
Said another way, when the magnitude of $\vec{q}$ scales, the angle of 
$\vec{K}_2 = (\vec{K}_1+\vec{q})$ must change.  
However, when we choose to work in representation (\ref{eq:K1K2Scheme}),
all momenta are fermionic which forces the wrong type of rescaling on the boson field.

Thus, we cannot adopt representation (\ref{eq:K1K2Scheme}) because, first, 
the $\Theta$-function is not form-invariant, and second, it forces a non-homogeneous
coordinate transformation on the boson field.

How might we try to remedy these two problems?
We could attempt the same strategy that worked in the pure fermion case
where we restricted our consideration to $\mathcal{K}_4 = 0$; see 
equations (\ref{eq:forward}) - (\ref{eq:cooper}).  However, when both bosons and fermions are present
this tactic is bound to fail.  We already found this in section~\ref{sec:bosons+fermions}
where we considered the limit $\mathcal{K}_2 = 0$ using representation (\ref{eq:KQScheme}).
Here, the analogous restriction is $\vec{ \mathcal{Q} } = 0$.  Under these circumstances, equivalent
to $\vec{K}_1 = \vec{K}_2$, the $\Theta$-function is trivially invariant.  However, the boson
loses its field character with $\phi(0)$ not scaling at all.  

Let us attempt a different remedy by relaxing the restriction slightly
and consider $\hat{K}_1 = \hat{K}_2$.  This is equivalent to
$\mathcal{Q} = | k_1 - k_2|$.  Here, the $\Theta$-function will not be
form invariant because it transforms to
\begin{eqnarray}
	\Theta(\Lambda - s^{1/z}|k_1 - k_2|)
	&=&
	\Theta(\Lambda - s^{(1-z)/z}|k_1^{\prime} - k_2^{\prime}|)	\nonumber
\end{eqnarray}
Undaunted, we make the further restriction to $z=1$, in which case the $\Theta$-function
is form-invariant, and now less trivially so.  Unfortunately, we still have the problem that
the boson field scales unnaturally: 
$\phi(|\vec{ \mathcal{Q} }|,\theta_q,\varphi_q) = \phi(|k_1-k_2|, \text{const},\text{const})$.
\begin{figure}[htbp]
   \centering
   \includegraphics[width=3in]{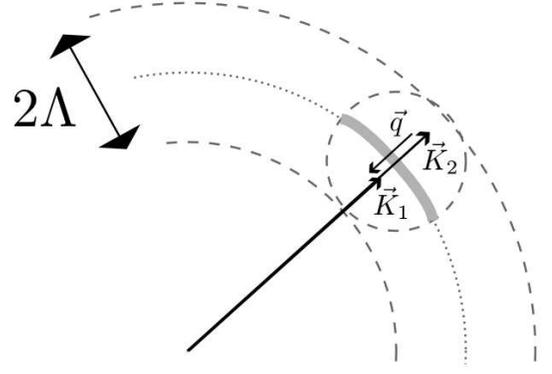}
   \caption[Kinematic constraints in the $K_1K_2$-scheme]
   {Under the restriction $\hat{K}_1 = \hat{K}_2$, the boson momentum  $\vec{q}$ can still
   have a non-zero modulus, but it loses its
   angular freedom and is forced to point exactly normal to the Fermi surface.
   This constraint does not allow the boson field $\phi$ to transform in a homogeneous
   fashion.  
   }
   \label{fig:K1K2PhaseSpace}
\end{figure}
As show in figure~\ref{fig:K1K2PhaseSpace}, the boson momentum vector, which is defined as the
vector joining the tips of $\vec{K}_1$ and $\vec{K}_2$, lies directly parallel to $\hat{K}_1 = \hat{K}_2$.
When we force $\hat{K}_1 = \hat{K}_2$, the boson momentum loses its angular freedom
and thus no longer scales homogeneously.

One might wonder why we are being so strict about the form of the field transformation when it seems like the
other scheme in equation (\ref{eq:KQScheme})
\begin{eqnarray}
\label{eq:KQScaling}
	\int^{\Lambda} d^dK d^dq\left[ \bar{\psi}_{\vec{K}+\vec{q}}\psi_{\vec{K}} \phi_{\vec{q}}\,g(\vec{K},\vec{q})
	\Theta(\Lambda - |\mathcal{K}_2|)\right]
\end{eqnarray}
also violates this principle.  In fact, the fermion is not required to be a homogeneous function
of momentum anyway.  All that we need from equation (\ref{eq:FermionScaling}) is
\begin{eqnarray}
	\psi(K_F+ s^{-1}k^{\prime} )
	&=&
	s^{3/2} \psi^{\prime}(K_F+ k^{\prime} )
\end{eqnarray}
The incoming fermion in equation (\ref{eq:KQScaling}) is clearly of this form, 
whereas the outgoing fermion can be written:
\begin{eqnarray}
	\bar{\psi}(|\vec{K}+\vec{q}|) 
	&\approx& \bar{\psi}(K_F + s^{-1}k^{\prime}+s^{-1}q^{\prime} \cos\theta_{Kq})
\end{eqnarray}
In this form, we know this expression is equivalent to:
\begin{eqnarray}
	&& \bar{\psi}(K_F + s^{-1}k^{\prime}+s^{-1}q^{\prime}\cos\theta_{Kq})
\nonumber \\
	&=& s^{3/2}\bar{\psi}^{\prime}(K_F + k^{\prime}+q^{\prime}\cos\theta_{Kq}) 
\end{eqnarray}
Thus, both fermions in equation (\ref{eq:KQScaling}) transform as 
equation (\ref{eq:FermionScaling}).
Finally, $\phi(\vec{q},i\omega)$ obviously scales according to equation (\ref{eq:phiScaling}) [and (\ref{eq:BosonScaling})].
Therefore,
in equation (\ref{eq:KQScaling}) we know how all fields transform under
equations (\ref{eq:enDim}), 
so representation (\ref{eq:KQScheme}) suffers from none of the 
shortcomings we identified for representation (\ref{eq:K1K2Scheme}).

With this new understanding, we should also check that in the pure 
fermion problem the field $\psi(\mathcal{K}_4)$ transforms in a consistent manner.  
To see this, we need to keep a few more higher order terms than what we showed earlier.
\begin{eqnarray}
	|\vec{K}_4|^2 
	&=& K_1^2+K_2^2+K_3^2 
	+ 2\vec{K}_1\cdot\vec{K}_2
	-2\vec{K}_1\cdot\vec{K}_3 \nonumber \\
&&	-2\vec{K}_2\cdot\vec{K}_3 \nonumber \\
	&\approx& 2K_F(k_1+k_2+k_3+3K_F/2) \nonumber \\
&&	+ 2K_F\hat{K}_1\cdot\hat{K}_2(k_1+k_2+K_F) \nonumber\\
&&	- 2K_F\hat{K}_1\cdot\hat{K}_3(k_1+k_3+K_F)\nonumber \\
&&	- 2K_F\hat{K}_2\cdot\hat{K}_3(k_2+k_3+K_F) 
\nonumber \\
	&=& 2K_F \Big\{
	K_F(\hat{K}_1\cdot\hat{K}_2-\hat{K}_1\cdot\hat{K}_3-\hat{K}_2\cdot\hat{K}_3 + 3/2) \nonumber \\
&&	+ k_1[1+\hat{K}_1\cdot(\hat{K}_2 - \hat{K}_3)] \nonumber \\
&&	+ k_2[1+\hat{K}_2\cdot(\hat{K}_1 - \hat{K}_3)] \nonumber \\
&&	+ k_3[1-\hat{K}_3\cdot(\hat{K}_1 + \hat{K}_2)] \Big\} 
\nonumber \\
	&=& 2K_F \Big\{
	K_F|\vec{\Delta}|^2/2\nonumber \\
&&	+ k_1[1+\hat{K}_1\cdot(\hat{K}_2 - \hat{K}_3)] \nonumber \\
&&	+ k_2[1+\hat{K}_2\cdot(\hat{K}_1 - \hat{K}_3)] \nonumber \\
&&	+ k_3[1-\hat{K}_3\cdot(\hat{K}_1 + \hat{K}_2)] \Big\} 
\end{eqnarray}
where we used (\ref{eq:modDelta}).
In the special case where $\hat{K}_1 = \hat{K}_3$, corresponding
to forward scattering, we have
\begin{eqnarray}
	|\vec{K}_4|
	&\approx&
	K_F + k_2+(k_1 - k_3)\hat{K}_1\cdot\hat{K}_2
\end{eqnarray}
This shows that
\begin{eqnarray}
	\psi(|\vec{K}_4|) 
	&=& \psi(K_F +s^{-1} k_2^{\prime}+s^{-1}(k_1^{\prime} - k_3^{\prime})\hat{K}_1\cdot\hat{K}_2) \nonumber\\
\end{eqnarray}
which is precisely the scaling form appropriate for a fermion in equation (\ref{eq:FermionScaling}).
In the same way, it is easy to show that the fermion scales appropriately for the cases
$\hat{K}_2 = \hat{K}_3$ and
$\hat{K}_1 = -\hat{K}_2$.
In these cases we have
\begin{eqnarray}
	|\vec{K}_4|
	&\approx&
	K_F + k_1+(k_2 - k_3)\hat{K}_1\cdot\hat{K}_3 \\
	|\vec{K}_4|
	&\approx&
	K_F + k_3+(k_2 - k_1)\hat{K}_1\cdot\hat{K}_3
\end{eqnarray}
respectively.  Thus, all the results of Shankar remain valid.

Finally, it may at first seem puzzling that the scaling of the constraint in
equation (\ref{eq:K1K2Constraint}) is so problematic since we were 
able to find a simple solution in the pure fermion problem
involving $\mathcal{S}_4^f$.  There, the constraint involved 
$\mathcal{K}_4 = |\vec{K}_3 - \vec{K}_2 - \vec{K}_1 | - K_F$, 
which measures a deviation from the Fermi surface.  
However, in the representation of the boson-fermion coupling in
equation (\ref{eq:K1K2Scheme}), the constraint involves 
$\vec{\mathcal{Q}} = \vec{K}_2 - \vec{K}_1$ which is {\it not} a deviation from the
Fermi surface and as written, can take any value between 0 and $2K_F$; 
see equation (\ref{eq:approxQ}).

To summarize, 
$\psi(\mathcal{K}_4)$ scales like a fermion,
$\psi(\mathcal{K}_2)$ scales like a fermion,
but
$\phi(\vec{\mathcal{Q} })$ does {\it not} scale like a boson.
We therefore cannot use equation (\ref{eq:K1K2Scheme})
to represent the boson-fermion coupling because we do not
have knowledge of the boson field scaling under
such a coordinate transformation.

\section{The Patching Scheme}
\label{sec:patching}
When $z \neq 1$, the scheme we developed in section~\ref{sec:bosons+fermions} no longer works.  
The problem is that under mode elimination
and rescaling, the constraint function changes its form.
\begin{eqnarray}
	\Theta(\Lambda/s - |\vec{\mathcal{K}_2}|) 
		&\approx& \Theta(\Lambda/s - | k+q\cos\theta_{Kq} | ) \nonumber \\
		&=& \Theta(\Lambda - s| k+q\cos\theta_{Kq} | ) \nonumber \\
		&=& \Theta(\Lambda - | k^{\prime}+s^{(z-1)/z} q^{\prime}\cos\theta_{Kq} | ) \nonumber \\
\end{eqnarray}
where $k^{\prime} = sk$ and $q^{\prime} = s^{1/z}q$.  
When $z \neq 1$, we cannot reliably determine the flow
of the coupling because the structure of the interaction itself has changed under 
this RG transformation.  
This is the same dilemma encountered in the pure fermion problem in equation
(\ref{eq:fermionConstraint}).
Also, notice that writing $\mathcal{S}_3^{bf}$ in terms of an integral over
$\vec{K}_1$ and $\vec{K}_2$, rather than $\vec{K}$ and $\vec{q}$, will not cure the problem.
In the previous section we explained why this is the case, even for $z =1$.  

For these reasons, when $z \neq 1$ we adopt a different method where we scale toward a specific
point on the Fermi surface.  Although the details differ, this is similar in spirit 
to some previous work on the renormalization of the
gauge-spinon problem~\cite{Polchinski1994, Nayak1994, Altshuler1994, Onoda1995}.

In $d=2$, consider the annulus in momentum space defined by $-\Lambda < k < \Lambda$.  Now subdivide the annulus into $N_{\Lambda}$ regions of angular size $\Delta\theta$: $N_{\Lambda} \Delta\theta = 2\pi \implies \Delta\theta = 2\pi\Lambda/K_F$.  Each patch will be approximately of size $\sim \Lambda^2$.  The same idea is easily generalized to $d>2$. This should be familiar from multidimensional bosonization~\cite{Kopietz1997} and 
functional RG~\cite{Furukawa1998, Zanchi2000a, Zanchi2000b, Honerkamp2001}; we refer the reader to those papers for further details.

The momentum integral for the quadratic part of the fermionic action, $\mathcal{S}_2^f$,  can now be decomposed into a sum over $N_{\Lambda}$ identical patches:
\begin{eqnarray}
	\mathcal{S}_2^f
		&=& \int^{\Lambda} d^dKd\epsilon \bar{\psi}(i\epsilon - \xi_{\vec{K}})\psi \nonumber \\
		&=& \sum_{p=1}^{N_{\Lambda}} \int^{\Lambda} d^dk_p d\epsilon_p \bar{\psi}_p(i\epsilon_p - \xi_{\vec{k}_p})\psi_p
\end{eqnarray}
Here, $\vec{k}_p = (\vec{k}_{p,\perp}, k_{p,\parallel})$ is a local coordinate within each patch which has components parallel and perpendicular to some reference frame.  We define this special local reference direction to be the normal vector to the Fermi surface at the patch origin.  Thus, $\vec{k}_{\perp}$ is tangent to the Fermi surface at the patch origin.  Within each patch, functions of momentum can be expanded around the patch origin.
Consider, for example, the patch centered at $\vec{K} = (0,K_F)$ which we will label as patch $p=1$, and where we have specialized to $d=2$ for concreteness.  
Near this point, the dispersion of a perfectly parabolic band 
can be expressed in terms of local patch coordinates as follows.
\begin{eqnarray}
	\xi_{\vec{K} \approx (0,K_F)}
		&\approx& \frac{1}{2m}\Big[ (K_x^2+K_y^2)|_{(0,K_F)} \nonumber \\
		&&+2K_x|_{(0,K_F)}(K_x-0) \nonumber\\
		&&+ 2K_y|_{(0,K_F)} (K_y-K_F)  \nonumber \\
		&&+\frac{1}{2m} \frac{1}{2}2|_{(0,K_F)}(K_x-0)^2 \nonumber\\
		&&+\frac{1}{2m} \frac{1}{2}2|_{(0,K_F)} (K_y-K_F)^2 \Big]  - \frac{K_F^2}{2m} \nonumber \\
		&\approx& v_F k_{1,\parallel} + \frac{k_{1,\perp}^2}{2m} \nonumber \\
		&=& v_F k_{1, \parallel} + \frac{v_Fk_{1,\perp}^2}{2K_F} \nonumber \\
		&\equiv& v_F k_{1, \parallel} + a k_{1,\perp}^2
\end{eqnarray}
where for this particular patch, $k_{1,\parallel} \equiv K_y - K_F$, 
and $k_{1,\perp} \equiv K_x$.
We have also defined $v_F \equiv K_F/m$ and $a \equiv v_F/(2K_F) = 1/(2m)$.  
As a sum over all the patches that enclose the Fermi surface, 
the quadratic part of the action can now be written
\begin{eqnarray}
		\mathcal{S}_2^f &=&
		\sum_{p=1}^{N_{\Lambda}} \int^{\Lambda} d^dk_p d\epsilon_p 
		\bar{\psi}_p\left(
		i\epsilon_p - v_F k_{p,\parallel} - a\vec{k}_{p,\perp}^2
		\right)\psi_p \nonumber
\end{eqnarray}
Note that the concepts of parallel and perpendicular only make sense with respect to a perfectly flat surface, or the normal to a specific point on a curved surface.  We take this specific point to be the center of the patch.
Momentum components in the same direction as the vector normal to the Fermi surface at the patch origin
are considered ``parallel," whereas momenta tangent to the Fermi surface are labeled ``perpendicular.''
We caution that different conventions exist in the literature for what is deemed parallel or perpendincular.  
We adopt the convention of reference~\cite{Altshuler1994}.

Within each patch, the momentum integral is limited to a box of dimension $\Lambda$ in every direction.
For example, in $d=2$ this means
\begin{eqnarray}
	\int^{\Lambda} d^2 k \equiv \int_{-\Lambda}^{\Lambda} dk_{\parallel}\int_{-\Lambda}^{\Lambda} dk_{\perp}
\end{eqnarray}
We have dropped the patch indices since we assume all patches are identical
in the sense that variables scale in the same manner in every patch.
In the absence of van Hove singularities and
nesting instabilities, this is a reasonable assumption
~\footnote{Actually, van Hove singularities may lead
to the dominance of some patches over others.
Such an idea has been explored, for example,
by \cite{Furukawa1998}, where a 2-patch model was employed.
This has been generalized to N-patches in the
functional RG literature.}.

\begin{figure}[htbp]
   \centering
   \includegraphics[width=3in]{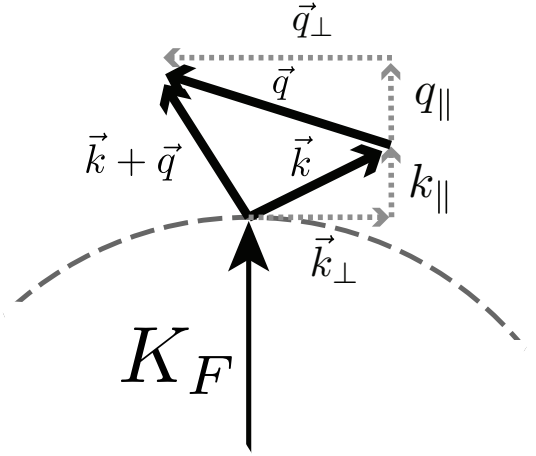}
   \caption{Local coordinate system in patch $p=1$ whose center is located at $\vec{K} = (0,K_F)$.
   Bosonic and fermionic momenta now scale identically}
   \label{fig:patch}
\end{figure}

Within this patching formalism, and when we consider only $\vec{q} \approx 0$
so that the entire boson phase space can be restricted to a single patch, 
the quadratic part of the bosonic action
can be written in straightforward fashion.
To be concrete, consider $z=3$:
\begin{eqnarray}
		\mathcal{S}_2^b&=&
		\int^{\Lambda} d^dq d\omega\;
		\phi^*
		\left( \vec{q}_{\perp}^2+ q_{\parallel}^2
		+\frac{\gamma \omega}{\sqrt{\vec{q}_{\perp}^2+q_{\parallel}^2} } \right)
		\phi 
		\nonumber
\end{eqnarray}
Within each patch, bosonic momenta $\vec{q}$ and fermionic momenta $\vec{k}$
are all measured with respect to the same single point, the patch origin.
Consequently, bosonic and fermionic momenta scale the same way, that is
\begin{eqnarray}
	\left[\vec{k}_{\perp}\right] &=& \left[\vec{q}_{\perp}\right] \\
	\left[k_{\parallel}\right] &=& \left[q_{\parallel}\right] \label{eq:kqparallel}
\end{eqnarray}
See figure~\ref{fig:patch}.  
Whether we label momenta by $\vec{k}$ or $\vec{q}$ is thus
immaterial since they scale identically; this is in stark contrast
to the scheme developed in section~\ref{sec:bosons+fermions}
for the case $z = 1$.

Of course, the possibility exists that $[k_{\parallel}] 
\neq [\vec{k}_{\perp}]$.  
In fact, we will now argue why they {\it cannot} be the same
when $z \neq 1$.

The fixed point is defined by constructing the scaling scheme so that
the quadratic part of the action, $\mathcal{S}_2^b+\mathcal{S}_2^f$, 
is scale invariant.
Scale invariance of $\mathcal{S}_2^f$ requires
\begin{eqnarray}
	[\epsilon] &=& [v_F]+ [k_{\parallel}] = [a]+2[\vec{k}_{\perp}]
\end{eqnarray}
while scale invariance of $\mathcal{S}_2^b$ necessitates
\begin{eqnarray}
	\left[\omega\right] + \left[\gamma\right]
	&=&
	\left[ \left( \vec{q}_{\perp}^2+ q_{\parallel}^2 \right)^{z/2} \right]
\end{eqnarray}
Next, we observe that since we want to scale bosons and fermions
simultaneously, it is sensible to give them equal scaling dimensions
which we denote by
\begin{eqnarray}
	\left[ E \right] 	&\equiv&	[\omega] = [\epsilon]
\end{eqnarray}
The conditions from the quadratic parts of the action now become
\begin{eqnarray}
	\left[ E \right] &=& [v_F]+ [k_{\parallel}] = [a]+2[\vec{k}_{\perp}] \\
	\left[ E \right] + \left[ \gamma \right] &=& \left[ \left( \vec{q}_{\perp}^2+ q_{\parallel}^2 \right)^{z/2} \right] \label{eq:S2bScaling}
\end{eqnarray}
At this point we demand that the dispersion relations of low-energy
excitations be preserved under scaling.
For fermions near the Fermi surface, energy must be a linear function of momentum.
We thus set $[v_F]=0$ to establish
\begin{eqnarray}
	\left[ E \right] &=& [k_{\parallel}] = [a]+2[\vec{k}_{\perp}] 
\label{eq:ParallelScaliesWithE}
\end{eqnarray}
This is furthermore justified by the fact that 
$\frac{a k_{\perp}^2 }{v_F k_{\parallel}} =  \frac{k_{\perp}^2}{K_F k_{\parallel}} \ll 1$
provided $k_{\perp, \parallel} \leq \Lambda \ll K_F$.
The ``curvature'' term is a small correction.
Thus, parallel momenta scale like energy.

We now use (\ref{eq:ParallelScaliesWithE}) and 
(\ref{eq:kqparallel}) in equation (\ref{eq:S2bScaling}) to determine
the dimension of the perpendicular momentum:
\begin{eqnarray}
	\left[ \gamma E \right] &=& \left[ \left( \vec{q}_{\perp}^2
+ v_F^2E^2 \right)^{z/2} \right] 
\nonumber \\
\implies 
	\left[ \vec{q}_{\perp} \right] &=& \left[ \sqrt{(\gamma E)^{2/z} - v_F^2E^2} \right]
\end{eqnarray}
In the infrared limit, this becomes
\begin{eqnarray}
	\left[ \vec{q}_{\perp} \right] &=& \left( \left[ E \right] + \left[ \gamma \right] \right)/ z
\end{eqnarray}
because $E^{2/z} > E^2$ when $z>1$.
To preserve the bosonic dispersion ({\it i.e.} $\omega \sim q^z$), 
we set $\left[ \gamma \right] = 0$, obtaining
\begin{eqnarray}
	\left[ \vec{q}_{\perp} \right] &=& \left[ E \right] / z
\end{eqnarray}
Now we plug this result into equation (\ref{eq:ParallelScaliesWithE})
to find the dimension of $a$:
\begin{eqnarray}
	\left[ E \right] &=& [a]+2 \left[ E \right]/z 
\nonumber \\
\implies
	\left[ a \right] &=& (1-2/z) \left[ E \right]
\end{eqnarray}
Finally, we are free to choose the value of $\left[ E\right]$,
which we set equal to unity for convenience;
any other value will only induce the same multiplicative
prefactor on all dimensions, but relative dimensions
will be unaffected.  To summarize,
\begin{eqnarray}
	\left[ E \right] &=& \left[ \epsilon \right] = \left[ \omega \right] = 1 \label{eq:energyScaling}
\nonumber \\
	\left[ k_{\parallel} \right] &=& \left[ q_{\parallel} \right] = \left[E\right] = 1 \label{eq:samePara} 
\nonumber \\
	\left[ \vec{k}_{\perp} \right] &=& \left[ \vec{q}_{\perp} \right] = \left[E\right]/z = 1/z \label{eq:samePerp} 
\nonumber \\
	\left[ a \right] &=& 1-2/z 
\nonumber \\
	\left[ v_F \right] &=& 0 
\nonumber \\
	\left[ \gamma \right] &=& 0
\label{eq:patchingDimensions}
\end{eqnarray}
Note that for $z=3$ the dimension of $\left[ \vec{k}_{\perp} \right]$
appears to suggest that the fermionic bandstructure changes under scaling.
This is an illusion since the parameter $a$, which is a measure of the curvature, is
allowed to scale in order to precisely compensate the scaling of $\vec{k}_{\perp}$, thus ensuring that the
band remains invariant.

Plugging these values into the quadratic action yields the dimensions of the fields:
\begin{eqnarray}
	\left[ \psi \right] &=&  -\frac{3z+d-1}{2z} \label{eq:psiDimPatch}\\
	\left[ \phi \right] &=&  -\frac{2z+d+1}{2z}\label{eq:phiDimPatch}
\end{eqnarray}

We now have enough information to determine the dimension
of the boson-fermion coupling.
\begin{eqnarray}
	\mathcal{S}_3^{bf} 
	&=& 
	\sum_{\text{patches} } 
	\int 
	d^dk_1 d^dk_2 d^dq 
	d\epsilon_2 d\epsilon_1 d\omega \nonumber \\
	&&\times g(\vec{k}_2,\vec{k}_1,\vec{q},\epsilon_2,\epsilon_1,\omega) \;
	 \bar{\psi}(2) \psi(1) \phi(\vec{q},\omega) \nonumber \\
	&&\times 
	\delta^{(d-1)}( \vec{k}_{2,\perp}  - \vec{k}_{1,\perp} - \vec{q}_{\perp} )
	\delta( k_{2,\parallel}  - k_{1, \parallel} - q_{\parallel} ) \nonumber \\
	&&\times \delta(\epsilon_2 - \epsilon_1 - \omega) \nonumber \\
	&&\times
	\Theta(\Lambda-|\vec{k}_{2,\perp}|)
	\Theta(\Lambda-|k_{2,\parallel}|) \nonumber \\
	&&\times \Theta(\Lambda-|\vec{k}_{1,\perp}|)
	\Theta(\Lambda-|k_{1,\parallel}|) \nonumber \\
	&&\times \Theta(\Lambda-|\vec{q}_{\perp}|) 
	\Theta(\Lambda-|q_{\parallel}|)
\end{eqnarray}
Note that this is slightly less general than could be the case.
We have restricted our consideration to nearly forward scattering
processes which means that $\vec{q} \approx 0$ or,
equivalently, $\vec{k}_1$ and $\vec{k}_2$
belong to the same patch.  Inter-patch processes,
such as the BCS instability, are not captured.

Since we are scaling toward a single point, momentum and energy
conserving delta functions and cutoff constraints factorize nicely.
Integrating against the delta functions yields
\begin{eqnarray}
	\mathcal{S}_3^{bf} 
	&=& 
	\sum_{\text{patches} } 
	\int^{\Lambda}
	d^dk d^dq 
	d\epsilon  d\omega \nonumber \\
	&&\times g(\vec{k}, \vec{q},\epsilon, \omega) \;
	 \bar{\psi}(\vec{k}+\vec{q},\epsilon+\omega) 
	 \psi(\vec{k},\epsilon)
	 \phi(\vec{q},\omega) \nonumber \\
	&&\times
	\Theta(\Lambda-|\vec{k}_{\perp}+\vec{q}_{\perp}|)
	\Theta(\Lambda-|k_{\parallel}+q_{\parallel}| ) 
\end{eqnarray}
where we have placed some of the constraints in the limits of integration.
Unlike what happened in section~\ref{sec:bosons+fermions},
there is no difference in eliminating boson or fermionic variables
due to equations (\ref{eq:patchingDimensions}).
Here it is arbitrary whether we call momentum
$\vec{k}$ or $\vec{q}$ since in the patching scheme they scale the same way.
Additionally, the factorization of parallel and perpendicular
components of momenta means the arguments of the fields scale in a straightforward
fashion.
Indeed, after mode elimination and rescaling we find:
\begin{eqnarray}
	\mathcal{S}_3^{bf} 
	&=& 
	g
	\sum_{\text{patches} } 
	\int^{\Lambda}
	s^{-1}s^{-(d-1)/z}
	d^dk^{\prime} 
	s^{-1}s^{-(d-1)/z}
	d^dq^{\prime}  \nonumber \\
	&&\times 
	s^{-1}d\epsilon^{\prime}  s^{-1}d\omega^{\prime}
	s^{(3z+d-1)/(2z)}
	 \bar{\psi}^{\prime} \nonumber \\
	&&\times s^{(3z+d-1)/(2z)}
	 \psi^{\prime}
	s^{(2z+d+1)/(2z)}
	 \phi^{\prime} \nonumber \\
	&&\times
	\Theta(\Lambda/s^{1/z}-s^{-1/z}|\vec{k}_{\perp}^{\prime}+\vec{q}_{\perp}^{\prime}|)
	\Theta(\Lambda/s-s^{-1}|k_{\parallel}^{\prime}+q_{\parallel}^{\prime}| ) \nonumber \\
	&=& 
	s^{(3-d)/(2z)}g
	\sum_{\text{patches} } 
	\int^{\Lambda}
	d^dk^{\prime} 
	d^dq^{\prime} 
	d\epsilon^{\prime} d\omega^{\prime}
	 \bar{\psi}^{\prime}
	 \psi^{\prime} 
	 \phi^{\prime} \nonumber \\
	&&\times
	\Theta(\Lambda-|\vec{k}_{\perp}^{\prime}+\vec{q}_{\perp}^{\prime}|)
	\Theta(\Lambda-|k_{\parallel}^{\prime}+q_{\parallel}^{\prime}| ) 
\end{eqnarray}
where we have Taylor expanded $g$ and kept
the most relevant (constant) piece.
In this patching scheme, the constraints and fields
transform in a simple way, so we can simply read
off the dimension of the coupling.
\begin{eqnarray}
	\left[ g \right] &=& \frac{3-d}{2z}
\label{eq:gDim}
\end{eqnarray}
The relevance or irrelevance of this coupling is in some sense
arbitrary outside the context of a specific physical problem.  
The value of $[g]$ depends crucially
on the dimensions $[\phi]$ and $[\psi]$, and these will be determined
by the problem under consideration.  For example, 
in the context of magnetic phases of the Kondo lattice,
see~\cite{Yamamoto2007, Yamamoto2008, Yamamoto2008b}.

Several important comments are now in order:

\begin{itemize}
	\item The result in equation (\ref{eq:gDim}) is identical to (\ref{eq:gDimzEqualsOne})  when $z=1$.
	Therefore, the patching scheme developed in this section yields an
	answer equivalent to the extension of Shankar's scheme presented in
	section \ref{sec:bosons+fermions}
	using global coordinates.  
	While the latter approach is perhaps more intuitive, it is not
	justifiable when $z \neq 1$.  On the other hand, the patching scheme
	requires same careful interpretation, as discussed below, but is consistent
	for any value of $z$.
		
	\item It is necessary to give the curvature parameter, $a$, a non-zero scaling dimension
	in order to compensate for the way that $k_{\perp}$ scales.  Rest assured, however, that
	$[a k_{\perp}^2] = [v_F k_{\parallel}] = [\epsilon]$ so that the fermion band is kept invariant.
	In this way, we do not need to scale the number of patches.
	
	\item It may seem as if the bosons have become anisotropic, but this is an illusion due to the nature
	of the local coordinates we have chosen.  Because of the sum over patches, we have included an
	equal weighting of $\vec{q}$ components in all directions, even though locally we only
	keep $\vec{q}_{\perp}$ within each patch.
	Of course,
	it does mean that in the low-energy limit bosons become locally tangent to the Fermi surface
	for fixed value of fermionic momentum $\vec{K}$.
	This is not surprising and was noticed long ago~\cite{Polchinski1994, Altshuler1994}.
	We even saw hints of this in section~\ref{sec:bosons+fermions}.  In that scaling scheme
	$k^{\prime} = sk$ and $q^{\prime} = s^{1/z}q$.  When $z>1$, the length of $|\vec{q}|$
	scales more slowly than the deviation from the Fermi surface, $k$.  As a result, in the low
	energy limit, the boson momentum will tend to lie tangent to the Fermi surface.
	
	\item In the patching formalism, the dimension of the boson field in equation
	(\ref{eq:phiDimPatch}) derives from
	\begin{equation}
		\quad\quad
		\phi^{\prime} 
		(q_{\parallel}^{\prime}, \vec{q}_{\perp}^{\prime}, i \omega^{\prime})
		=
		s^{ [\phi] } \phi
		(s^{-[q_{\parallel} ] } q_{\parallel}^{\prime}, s^{-[\vec{q}_{\perp} ] } \vec{q}_{\perp}^{\prime}, s^{-[\omega]} i \omega^{\prime})
		 \nonumber
	\end{equation}
	and similarly for the fermion field.  Once again this takes the form of a generalized homogeneous function,
	but is different from the type of scaling in equation (\ref{eq:BosonScaling}) or (\ref{eq:FermionScaling}).
	For more on generalized homogeneous functions, see~\cite{Hankey1972}.
	
	\item The form of the interaction we consider is limited to nearly forward scattering ($q \approx 0$)
	intra-patch processes.  Inter-patch processes are not captured, and this makes comparisons
	with the pure-fermion RG somewhat delicate.  Consider a four-fermion
	interaction with incoming momenta $\vec{K}_1$ and $\vec{K}_2$, and outgoing momenta
	$\vec{K}_3$ and $\vec{K}_4$.  The difference between incoming and outgoing momenta
	at the left vertex can be small, say $\vec{K}_3 - \vec{K}_1 \equiv \vec{q}_{\text{left vertex}} \approx0$.  
	This can
	match up with small momentum transfer on the right: 
	$\vec{K}_4 - \vec{K}_2 \equiv \vec{q}_{\text{right vertex}} \approx0$.
	However, this says nothing about the relationship between $\vec{K}_1$ and $\vec{K}_2$.
	Indeed, $\vec{K}_1$ and $\vec{K}_2$ can each
	independently take any value around the Fermi surface, {\it i.e.} $|\vec{K}_2 - \vec{K}_1|$
	can take any value between $0$ and $2K_F$.  Thus, ``forward scattering'' processes in a 
	boson-fermion formalism are not necessarily equivalent to ``forward scattering''
	processes in a four-fermion formalism.  The latter (four-fermion coupling) involves two patches, 
	whereas the former (boson-fermion coupling) involves only one patch.
	In other words, the dimension of $u_f$ is \textit{not} simply given by $[g^2]$.
	
	\item If we were to include self energy corrections into $\mathcal{S}_2^f$ and establish this
	as the new fixed point, the values of the dimension assignments would change, but the philosophy
	would be the same.  For example, in the gauge-spinon~\cite{Altshuler1994} 
	and ferromagnetic Kondo lattice systems~\cite{Yamamoto2008b},
	gapless overdamped $z=3$ bosons lead to a characteristic electron self energy
	$\Sigma(\epsilon) \sim \epsilon^{2/3}$ in $d=2$ and 
	$\Sigma(\epsilon) \sim -\epsilon\log\epsilon$ in $d=3$.  We can define the new fixed-point action
	with $\mathcal{S}_2^f = \int \bar{\psi}(\epsilon^{d/z} - v_F k_{\parallel} - a k_{\perp}^2)\psi$.
	Using the same philosophy defined in this section, we would assign 
\begin{eqnarray}
	\left[E\right] &=& \left[ \epsilon \right] = \left[ \omega \right] = z/d 
\nonumber \\
	\left[ k_{\parallel} \right] &=& \left[ q_{\parallel} \right] = \left[E\right] = 1
\nonumber \\
	\left[ \vec{k}_{\perp} \right] &=& \left[ \vec{q}_{\perp} \right] = \left[E\right]/z = 1/d 
\nonumber \\
	\left[ a \right] &=& 1-2/d 
\nonumber \\
	\left[ v_F \right] &=& 0 
\nonumber \\
	\left[ \gamma \right] &=& 0
\end{eqnarray}
	This also leads to a change in the dimensions of the fields and the couplings,
	but the methodology is no different than what has already been discussed above.
	See~\cite{Yamamoto2008b} for further discussion.

	\item There is some debate in the literature about how to properly scale
	the gauge-spinon model which corresponds to 
	$z=3$~\cite{Polchinski1994, Altshuler1994, Nayak1994, Onoda1995, Lee2008}.
	The consistent scaling scheme within the patching formalism we advocate
	here coincides with that of~\cite{Altshuler1994}.
\end{itemize}

\section{Conclusion}
\label{sec:conclusion}
This paper has developed an easy-to-use RG procedure for theories 
containing both bosons and fermions with a Fermi surface.  We reviewed the 
global coordinate approach to the fermionic RG as formulated by Shankar, showed how to generalize this
formalism to include bosons with dynamical exponent $z=1$, and explained why such
an approach will not work when $z \neq 1$.  We pointed out that a consistent 
scheme must ensure that the kinematic constraints, which result from the 
conservation of momentum and the effective field theory cutoffs, 
remain invariant to the RG transformation. In addition, field rescaling can only be properly 
identified in interaction terms when the coordinates of the field transform 
in a known way, as specified by the quadratic part of the action.  

We also showed that, for $z=1$, the same results arise within a patching
scheme. Here the momentum space near the Fermi surface is partitioned
into patches. For $z \neq 1$, the patching scheme represents the only
consistent RG approach to mixed fermion-boson systems.

Coupled boson and fermion problems arise in a variety of contexts. We have 
already mentioned the problems of itinerant magnets which have directly
motivated our work here, as well as the subject of gauge fields coupled to fermions. 
In addition, fermion-boson mixtures of cold atomic gases~\cite{Yang2008}
may provide another interesting setting for this work.
We hope the RG program 
described here will be useful for related problems in other settings as well.

\section{Acknowledgements}
We thank R. Shankar for useful discussions on the fermionic RG.
This work has been supported in part by the NSF Grant No. DMR-0706625,
the Robert A. Welch Foundation, and the W. M. Keck Foundation.
One of us (Q.S.) gratefully acknowledges the Aspen Center for Physics
for hospitality.


\begin{thebibliography}{30}
\expandafter\ifx\csname natexlab\endcsname\relax\def\natexlab#1{#1}\fi
\expandafter\ifx\csname bibnamefont\endcsname\relax
  \def\bibnamefont#1{#1}\fi
\expandafter\ifx\csname bibfnamefont\endcsname\relax
  \def\bibfnamefont#1{#1}\fi
\expandafter\ifx\csname citenamefont\endcsname\relax
  \def\citenamefont#1{#1}\fi
\expandafter\ifx\csname url\endcsname\relax
  \def\url#1{\texttt{#1}}\fi
\expandafter\ifx\csname urlprefix\endcsname\relax\def\urlprefix{URL }\fi
\providecommand{\bibinfo}[2]{#2}
\providecommand{\eprint}[2][]{\url{#2}}

\bibitem[{\citenamefont{Wilson and Kogut}(1974)}]{Wilson1974}
\bibinfo{author}{\bibfnamefont{K.~G.} \bibnamefont{Wilson}} \bibnamefont{and}
  \bibinfo{author}{\bibfnamefont{J.~B.} \bibnamefont{Kogut}},
  \bibinfo{journal}{Phys. Rep.} \textbf{\bibinfo{volume}{2}},
  \bibinfo{pages}{75} (\bibinfo{year}{1974}).

\bibitem[{\citenamefont{Hertz}(1976)}]{Hertz1976}
\bibinfo{author}{\bibfnamefont{J.~A.} \bibnamefont{Hertz}},
  \bibinfo{journal}{Phys. Rev. B} \textbf{\bibinfo{volume}{14}},
  \bibinfo{pages}{1165} (\bibinfo{year}{1976}).

\bibitem[{\citenamefont{Feldman and Trubowitz}(1990)}]{Feldman1990}
\bibinfo{author}{\bibfnamefont{J.}~\bibnamefont{Feldman}} \bibnamefont{and}
  \bibinfo{author}{\bibfnamefont{E.}~\bibnamefont{Trubowitz}},
  \bibinfo{journal}{Helv. Phys. Acta.} \textbf{\bibinfo{volume}{63}},
  \bibinfo{pages}{156} (\bibinfo{year}{1990}).

\bibitem[{\citenamefont{Benfatto and Gallavotti}(1990)}]{Benfatto1990}
\bibinfo{author}{\bibfnamefont{G.}~\bibnamefont{Benfatto}} \bibnamefont{and}
  \bibinfo{author}{\bibfnamefont{G.}~\bibnamefont{Gallavotti}},
  \bibinfo{journal}{Phys. Rev. B} \textbf{\bibinfo{volume}{42}},
  \bibinfo{pages}{9967} (\bibinfo{year}{1990}).

\bibitem[{\citenamefont{Shankar}(1991)}]{Shankar1991}
\bibinfo{author}{\bibfnamefont{R.}~\bibnamefont{Shankar}},
  \bibinfo{journal}{Physica A} \textbf{\bibinfo{volume}{177}},
  \bibinfo{pages}{530} (\bibinfo{year}{1991}).

\bibitem[{\citenamefont{Shankar}(1994)}]{Shankar1994}
\bibinfo{author}{\bibfnamefont{R.}~\bibnamefont{Shankar}},
  \bibinfo{journal}{Rev. Mod. Phys.} \textbf{\bibinfo{volume}{66}},
  \bibinfo{pages}{129} (\bibinfo{year}{1994}).

\bibitem[{\citenamefont{Vojta et~al.}(1997)\citenamefont{Vojta, Belitz,
  Narayanan, and Kirkpatrick}}]{Vojta1997}
\bibinfo{author}{\bibfnamefont{T.}~\bibnamefont{Vojta}},
  \bibinfo{author}{\bibfnamefont{D.}~\bibnamefont{Belitz}},
  \bibinfo{author}{\bibfnamefont{R.}~\bibnamefont{Narayanan}},
  \bibnamefont{and}
  \bibinfo{author}{\bibfnamefont{T.}~\bibnamefont{Kirkpatrick}},
  \bibinfo{journal}{Z. Phys. B: Condens. Matter}
  \textbf{\bibinfo{volume}{103}}, \bibinfo{pages}{451} (\bibinfo{year}{1997}).

\bibitem[{\citenamefont{Abanov and Chubukov}(2004)}]{Abanov2004}
\bibinfo{author}{\bibfnamefont{A.}~\bibnamefont{Abanov}} \bibnamefont{and}
  \bibinfo{author}{\bibfnamefont{A.~V.} \bibnamefont{Chubukov}},
  \bibinfo{journal}{Phys. Rev. Lett.} \textbf{\bibinfo{volume}{93}},
  \bibinfo{pages}{255702} (\bibinfo{year}{2004}).

\bibitem[{\citenamefont{Polchinski}(1994)}]{Polchinski1994}
\bibinfo{author}{\bibfnamefont{J.}~\bibnamefont{Polchinski}},
  \bibinfo{journal}{Nucl. Phys. B} \textbf{\bibinfo{volume}{422}},
  \bibinfo{pages}{617} (\bibinfo{year}{1994}).

\bibitem[{\citenamefont{Nayak and Wilczek}(1994)}]{Nayak1994}
\bibinfo{author}{\bibfnamefont{C.}~\bibnamefont{Nayak}} \bibnamefont{and}
  \bibinfo{author}{\bibfnamefont{F.}~\bibnamefont{Wilczek}},
  \bibinfo{journal}{Nucl. Phys. B} \textbf{\bibinfo{volume}{430}},
  \bibinfo{pages}{534} (\bibinfo{year}{1994}).

\bibitem[{\citenamefont{Altshuler et~al.}(1994)\citenamefont{Altshuler, Ioffe,
  and Millis}}]{Altshuler1994}
\bibinfo{author}{\bibfnamefont{B.}~\bibnamefont{Altshuler}},
  \bibinfo{author}{\bibfnamefont{L.}~\bibnamefont{Ioffe}}, \bibnamefont{and}
  \bibinfo{author}{\bibfnamefont{A.~J.} \bibnamefont{Millis}},
  \bibinfo{journal}{Phys. Rev. B} \textbf{\bibinfo{volume}{50}},
  \bibinfo{pages}{14048} (\bibinfo{year}{1994}).

\bibitem[{\citenamefont{Onoda et~al.}(1995)\citenamefont{Onoda, Ichinose, and
  Matsui}}]{Onoda1995}
\bibinfo{author}{\bibfnamefont{M.}~\bibnamefont{Onoda}},
  \bibinfo{author}{\bibfnamefont{I.}~\bibnamefont{Ichinose}}, \bibnamefont{and}
  \bibinfo{author}{\bibfnamefont{T.}~\bibnamefont{Matsui}},
  \bibinfo{journal}{Nucl. Phys. B} \textbf{\bibinfo{volume}{446}},
  \bibinfo{pages}{353} (\bibinfo{year}{1995}).

\bibitem[{\citenamefont{Holstein et~al.}(1973)\citenamefont{Holstein, Norton,
  and Pincus}}]{Holstein1973}
\bibinfo{author}{\bibfnamefont{T.}~\bibnamefont{Holstein}},
  \bibinfo{author}{\bibfnamefont{R.}~\bibnamefont{Norton}}, \bibnamefont{and}
  \bibinfo{author}{\bibfnamefont{P.}~\bibnamefont{Pincus}},
  \bibinfo{journal}{Phys. Rev. B} \textbf{\bibinfo{volume}{8}},
  \bibinfo{pages}{2649} (\bibinfo{year}{1973}).

\bibitem[{\citenamefont{Varma et~al.}(2002)\citenamefont{Varma, Nussinov, and
  Saarloos}}]{Varma2002}
\bibinfo{author}{\bibfnamefont{C.~M.} \bibnamefont{Varma}},
  \bibinfo{author}{\bibfnamefont{Z.}~\bibnamefont{Nussinov}}, \bibnamefont{and}
  \bibinfo{author}{\bibfnamefont{W.~v.} \bibnamefont{Saarloos}},
  \bibinfo{journal}{Phys. Rep.} \textbf{\bibinfo{volume}{361}},
  \bibinfo{pages}{267} (\bibinfo{year}{2002}).

\bibitem[{\citenamefont{Lee et~al.}(2006)\citenamefont{Lee, Nagaosa, and
  Wen}}]{Lee2006}
\bibinfo{author}{\bibfnamefont{P.}~\bibnamefont{Lee}},
  \bibinfo{author}{\bibfnamefont{N.}~\bibnamefont{Nagaosa}}, \bibnamefont{and}
  \bibinfo{author}{\bibfnamefont{X.}~\bibnamefont{Wen}}, \bibinfo{journal}{Rev.
  Mod. Phys.} \textbf{\bibinfo{volume}{78}}, \bibinfo{pages}{17}
  (\bibinfo{year}{2006}).

\bibitem[{\citenamefont{Furukawa et~al.}(1998)\citenamefont{Furukawa, Rice, and
  Salmhofer}}]{Furukawa1998}
\bibinfo{author}{\bibfnamefont{N.}~\bibnamefont{Furukawa}},
  \bibinfo{author}{\bibfnamefont{T.~M.} \bibnamefont{Rice}}, \bibnamefont{and}
  \bibinfo{author}{\bibfnamefont{M.}~\bibnamefont{Salmhofer}},
  \bibinfo{journal}{Phys. Rev. Lett.} \textbf{\bibinfo{volume}{81}},
  \bibinfo{pages}{3195} (\bibinfo{year}{1998}).

\bibitem[{\citenamefont{Zanchi and Schulz}(2000{\natexlab{a}})}]{Zanchi2000a}
\bibinfo{author}{\bibfnamefont{D.}~\bibnamefont{Zanchi}} \bibnamefont{and}
  \bibinfo{author}{\bibfnamefont{H.~J.} \bibnamefont{Schulz}},
  \bibinfo{journal}{Europhys. Lett.} \textbf{\bibinfo{volume}{44}},
  \bibinfo{pages}{235} (\bibinfo{year}{2000}{\natexlab{a}}).

\bibitem[{\citenamefont{Zanchi and Schulz}(2000{\natexlab{b}})}]{Zanchi2000b}
\bibinfo{author}{\bibfnamefont{D.}~\bibnamefont{Zanchi}} \bibnamefont{and}
  \bibinfo{author}{\bibfnamefont{H.~J.} \bibnamefont{Schulz}},
  \bibinfo{journal}{Phys. Rev. B} \textbf{\bibinfo{volume}{61}},
  \bibinfo{pages}{609} (\bibinfo{year}{2000}{\natexlab{b}}).

\bibitem[{\citenamefont{Honerkamp et~al.}(2001)\citenamefont{Honerkamp,
  Salmhofer, Furukawa, and Rice}}]{Honerkamp2001}
\bibinfo{author}{\bibfnamefont{C.}~\bibnamefont{Honerkamp}},
  \bibinfo{author}{\bibfnamefont{M.}~\bibnamefont{Salmhofer}},
  \bibinfo{author}{\bibfnamefont{N.}~\bibnamefont{Furukawa}}, \bibnamefont{and}
  \bibinfo{author}{\bibfnamefont{T.~M.} \bibnamefont{Rice}},
  \bibinfo{journal}{Phys. Rev. B} \textbf{\bibinfo{volume}{63}},
  \bibinfo{pages}{035109} (\bibinfo{year}{2001}).

\bibitem[{\citenamefont{Schutz et~al.}(2005)\citenamefont{Schutz, Bartosch, and
  Kopietz}}]{Schutz2005}
\bibinfo{author}{\bibfnamefont{F.}~\bibnamefont{Schutz}},
  \bibinfo{author}{\bibfnamefont{L.}~\bibnamefont{Bartosch}}, \bibnamefont{and}
  \bibinfo{author}{\bibfnamefont{P.}~\bibnamefont{Kopietz}},
  \bibinfo{journal}{Phys. Rev. B} \textbf{\bibinfo{volume}{72}},
  \bibinfo{pages}{035107} (\bibinfo{year}{2005}).

\bibitem[{\citenamefont{Tsai et~al.}(2005)\citenamefont{Tsai, Castro~Neto,
  Shankar, and Campbell}}]{Tsai2005}
\bibinfo{author}{\bibfnamefont{S.}~\bibnamefont{Tsai}},
  \bibinfo{author}{\bibfnamefont{A.}~\bibnamefont{Castro~Neto}},
  \bibinfo{author}{\bibfnamefont{R.}~\bibnamefont{Shankar}}, \bibnamefont{and}
  \bibinfo{author}{\bibfnamefont{D.}~\bibnamefont{Campbell}},
  \bibinfo{journal}{Phys. Rev. B} \textbf{\bibinfo{volume}{72}},
  \bibinfo{pages}{054531} (\bibinfo{year}{2005}).

\bibitem[{\citenamefont{Yamamoto and Si}(2007)}]{Yamamoto2007}
\bibinfo{author}{\bibfnamefont{S.~J.} \bibnamefont{Yamamoto}} \bibnamefont{and}
  \bibinfo{author}{\bibfnamefont{Q.}~\bibnamefont{Si}}, \bibinfo{journal}{Phys.
  Rev. Lett.} \textbf{\bibinfo{volume}{99}}, \bibinfo{pages}{016401}
  (\bibinfo{year}{2007}).

\bibitem[{\citenamefont{Yamamoto and Si}(2008)}]{Yamamoto2008}
\bibinfo{author}{\bibfnamefont{S.~J.} \bibnamefont{Yamamoto}} \bibnamefont{and}
  \bibinfo{author}{\bibfnamefont{Q.}~\bibnamefont{Si}},
  \bibinfo{journal}{Physica B} \textbf{\bibinfo{volume}{403}},
  \bibinfo{pages}{1414} (\bibinfo{year}{2008}).

\bibitem[{\citenamefont{Yamamoto and Si}()}]{Yamamoto2008b}
\bibinfo{author}{\bibfnamefont{S.~J.} \bibnamefont{Yamamoto}} \bibnamefont{and}
  \bibinfo{author}{\bibfnamefont{Q.}~\bibnamefont{Si}},
  \emph{\bibinfo{title}{Metallic ferromagnetism in the {K}ondo lattice}},
  \bibinfo{note}{{a}r{X}iv:0812.0819}.

\bibitem[{\citenamefont{Moriya}(1985)}]{Moriya1985}
\bibinfo{author}{\bibfnamefont{T.}~\bibnamefont{Moriya}},
  \emph{\bibinfo{title}{Spin Fluctuations in Itinerant Electron Magnetism}}
  (\bibinfo{publisher}{Springer}, \bibinfo{year}{1985}).

\bibitem[{\citenamefont{Millis}(1993)}]{Millis1993}
\bibinfo{author}{\bibfnamefont{A.~J.} \bibnamefont{Millis}},
  \bibinfo{journal}{Phys. Rev. B} \textbf{\bibinfo{volume}{48}},
  \bibinfo{pages}{7183} (\bibinfo{year}{1993}).

\bibitem[{\citenamefont{Kopietz}(1997)}]{Kopietz1997}
\bibinfo{author}{\bibfnamefont{P.}~\bibnamefont{Kopietz}},
  \emph{\bibinfo{title}{Bosonization of Interacting Fermions in Arbitrary
  Dimensions}} (\bibinfo{publisher}{Springer-Verlag Telos},
  \bibinfo{address}{Cambridge}, \bibinfo{year}{1997}).

\bibitem[{\citenamefont{Hankey and Stanley}(1972)}]{Hankey1972}
\bibinfo{author}{\bibfnamefont{A.}~\bibnamefont{Hankey}} \bibnamefont{and}
  \bibinfo{author}{\bibfnamefont{H.~E.} \bibnamefont{Stanley}},
  \bibinfo{journal}{Phys. Rev. B} \textbf{\bibinfo{volume}{6}},
  \bibinfo{pages}{3515} (\bibinfo{year}{1972}).

\bibitem[{\citenamefont{Lee}(2008)}]{Lee2008}
\bibinfo{author}{\bibfnamefont{S.~S.} \bibnamefont{Lee}},
  \bibinfo{journal}{Phys. Rev. B} \textbf{\bibinfo{volume}{78}},
  \bibinfo{pages}{085129} (\bibinfo{year}{2008}).

\bibitem[{\citenamefont{Yang}(2008)}]{Yang2008}
\bibinfo{author}{\bibfnamefont{K.}~\bibnamefont{Yang}}, \bibinfo{journal}{Phys.
  Rev. B} \textbf{\bibinfo{volume}{77}}, \bibinfo{pages}{085115}
  (\bibinfo{year}{2008}).

\end{thebibliography}

\end{document}